\pdfoutput=1
\RequirePackage{ifpdf}
\ifpdf % We are running pdfTeX in pdf mode
\documentclass[pdftex]{sigma}
\else
\documentclass{sigma}
\fi

\numberwithin{equation}{section}

\begin{document}

\allowdisplaybreaks

\newcommand{\arXivNumber}{1703.04472}

\renewcommand{\PaperNumber}{054}

\FirstPageHeading

\ShortArticleName{Topological Phase Transitions in a Molecular Hamiltonian}

\ArticleName{Topological Phase Transition in a Molecular\\ Hamiltonian with Symmetry and Pseudo-Symmetry,\\ Studied through Quantum, Semi-Quantum\\ and Classical Models}

\Author{Guillaume DHONT~$^\dag$, Toshihiro IWAI~$^\ddag$ and Boris ZHILINSKII~$^\dag$}

\AuthorNameForHeading{G.~Dhont, T.~Iwai and B.~Zhilinski\'{\i}}

\Address{$^\dag$~Universit\'e du Littoral C\^ote d'Opale, Laboratoire de Physico-Chimie de l'Atmosph\`ere,\\
 \hphantom{$^\dag$}~189A Avenue Maurice Schumann, 59140 Dunkerque, France}
\EmailD{\href{mailto:guillaume.dhont@univ-littoral.fr}{guillaume.dhont@univ-littoral.fr}, \href{zhilin@univ-littoral.fr}{zhilin@univ-littoral.fr}}

\Address{$^\ddag$~Kyoto University, 606-8501 Kyoto, Japan}
\EmailD{\href{mailto:iwai.toshihiro.63u@st.kyoto-u.ac.jp}{iwai.toshihiro.63u@st.kyoto-u.ac.jp}}

\ArticleDates{Received March 14, 2017, in f\/inal form July 04, 2017; Published online July 13, 2017}

\Abstract{The redistribution of energy levels between energy bands is studied for a family of simple ef\/fective Hamiltonians depending on
 one control parameter and possessing axial symmetry and energy-ref\/lection symmetry. Further study is made on the topological phase transition in the corresponding semi-quantum and completely classical models, and f\/inally the joint spectrum of the two commuting observables $(H=E,J_z)$ (also called the lattice of quantum states) is superposed on the image of the energy-momentum map for the classical model. Through these comparative analyses, mutual correspondence is demonstrated to exist among the redistribution of energy levels between energy bands for the quantum Hamiltonian, the modif\/ication of Chern numbers of eigenline bundles for the corresponding semi-quantum Hamiltonian, and the presence of Hamiltonian monodromy for the complete classical analog. In particular, as far as the band rearrangement is concerned, a f\/ine agreement is found between the redistribution of the energy levels described in terms of joint spectrum of energy and momentum in the full quantum model and the evolution of singularities of the energy-momentum map of the complete classical model. The topological phase transition observed in the present semi-quantum and the complete classical models are analogous to topological phase transitions of matter.}


\Classification{03G10; 15B57; 53C80; 81V55}

\section{Introduction}
The theoretical and experimental study of topological phases of matter is a hot topic of mathe\-matical and solid state physics of the last twenty-f\/ive years \cite{Avron2, Bernevig,Haldane, HasanKane, Kitaev, Ryu10,Ryu08, Thouless, Volovik}. Less attention is paid to the qualitative ef\/fects occurring for f\/inite particle quantum systems, such as atomic or molecular systems, under the variation of some control parameter. Owing to a~suf\/f\/iciently high density of states, the dynamical behavior in excited states of such a simple few-body system nevertheless mimics a behavior quite similar to the topological phase transitions seen in condensed phase matter \cite{IwaiZhTheorChemAcc,Iachello, Stransky, BIZPhysRep}. For a~conceptual unity among dif\/ferent f\/ields of physics, it should be recognized that common topological ideas are shared among molecular physics, topological insulators, superf\/luids, and particle physics although they are formulated in dif\/ferent words in their respective f\/ields, such as band rearrangement or energy level redistribution, gap closing, gap node, gapless excitation, contact of the conduction band and the valence band at a single point, Dirac points, etc. Further comments on this aspect will be given in the last section.

The topological origin of the qualitative ef\/fect of redistribution of quantum energy levels between energy bands for simple isolated molecular systems under the variation of a control parameter was already suggested in 1988 \cite{EurophysL}. Later, topological invariants such as Chern numbers and especially delta-Chern numbers calculated within the associated semi-quantum model \cite{FaurePRL, AnnPhysIZh, IwaiZhWallCross, I_Z} were demonstrated to be relevant to the band rearrangement for a number of concrete molecular examples. The correspondence between the redistribution phenomenon for a quantum Hamiltonian and the appearance of Hamiltonian monodromy for its full classical analog was equally studied~\cite{RevModPhys, PhysLettMono}.

For a better understanding of the reorganization of the energy bands and of the classif\/ication of topological phases of matter, it is of great use to take into account for dynamical molecular Hamiltonians additional symmetries such as the time-reversal and the particle-hole transformations, whose actions are represented in quantum mechanics by antiunitary transformations~\cite[Chapter~26]{WignerBook}. Manifestations of these discrete symmetries are dif\/ferent for the fermionic bands characteristic for the topological insulators and superconductors \cite{Zirnba} from one side and for the adiabatic energy bands formed in f\/inite particle systems due to adiabatic separation of fast and slow dynamic degrees of freedom from another side~\cite{OmriGat}. Molecular examples with time-reversal symmetry and without time-reversal symmetry were both studied and a~general statement concerning the implication of time-reversal invariance on the Chern numbers of adiabatic energy bands was recently formulated~\cite{OmriGat}. At the same time the analysis of the energy-ref\/lection symmetry for molecular problems has not been seriously studied, whereas this kind of generalized symmetry (or pseudo-symmetry), known also as particle-hole or charge conjugation symmetry, is typical for condensed matter studies \cite{Dunne2, DunneShifman}. Systems with such a symmetry belong to the $D$ class (one of the BdG classes) within a ten-fold way classif\/ication of the topological insulators and superconductors \cite{Zirnba, Bernevig, Ryu08}. In the present paper we study molecular ef\/fective Hamiltonians with a band structure resulting from the separation of dynamical variables into slow and fast subsystems, on the assumption of the axial symmetry and of the dynamical pseudo-symmetry manifesting as an \textit{energy-reflection symmetry}.

\begin{figure}[t]\centering
 \includegraphics[scale=0.90]{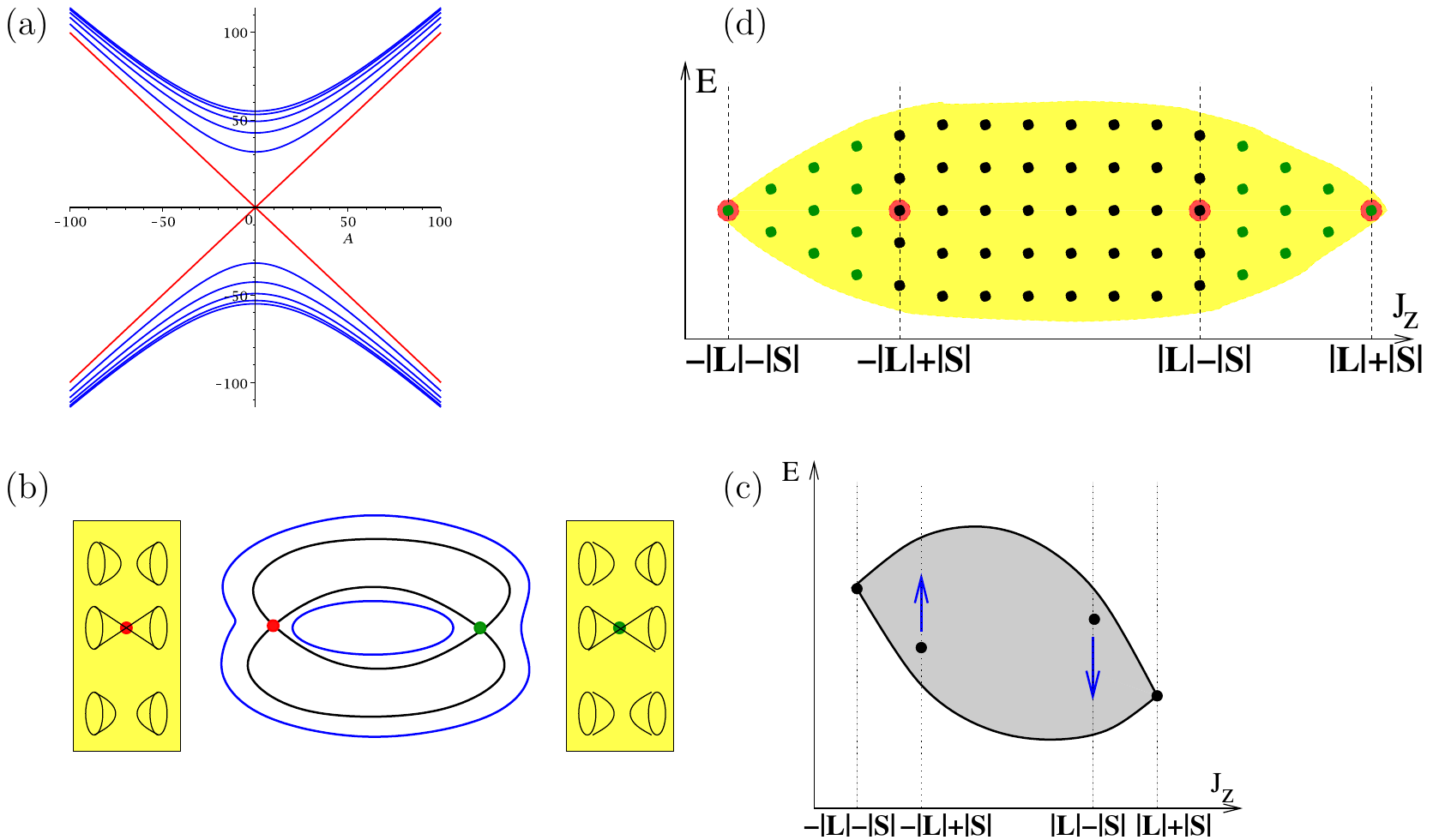}
\caption{Graphical summary of the four dif\/ferent descriptions of the same problem studied in the present paper. (a) Evolution of the quantum energy levels of Hamiltonian~(\ref{Hamiltonian_ref}) for $S=1/2$ as a function of control parameter $A$. (b) Schematic representation of the eigenvalues of the semi-quantum Hamiltonian for $S=1/2$ forming two conical intersection points viewed as a topological origin of the redistribution of energy levels between bands in the completely quantum version. (c) Image of the energy-momentum map for the classical Hamiltonian system with $L\gg S$, $A\sim0$. The four isolated critical values of the energy-momentum map are shown by f\/illed dots. For more details see Fig.~\ref{EnMomLim}. (d) Lattice of quantum states corresponding to the classical Hamiltonian $H$ with $L=5$, $S=2$, $A=0$ superposed on the image of the classical energy-momentum map. Quantum bulk and edge states are marked by black and green dots respectively, edge states being responsible for the redistribution of energy levels.}\label{ResumeGraf}
\end{figure}

The purpose of the present paper is to show that in the presence of the axial symmetry together with or without the energy-ref\/lection symmetry there exists a mutual correspondence among the redistribution of the energy levels between bands for the quantum Hamiltonian, the modif\/ication of the Chern numbers of the eigenline bundles for the corresponding semi-quantum Hamiltonian, and the presence of Hamiltonian monodromy for the complete classical analog together with f\/ine agreement between the evolution of the energy-momentum map and the redistribution of energy levels in the initial quantum system seen through the evolution of the joint spectrum of two commuting observables. Fig.~\ref{ResumeGraf} graphically summarizes the dif\/ferent aspects of the reorganization of the energy band structure and justif\/ies the qualif\/ication of this qualitative phenomenon as a~topological phase transition in an isolated molecular system.

The present paper is organized as follows: Section~\ref{S_Quantum} starts with the analysis of a simple quantum system possessing two energy bands whose internal structure is described by angular momentum variables. A description is given of the generic phenomenon of the redistribution of energy levels between energy bands for the two-band problem with the axial symmetry and an accent is put on the new features appearing in the presence of additional energy-ref\/lection symmetry. The ef\/fective Hamiltonian under study depends on one control parameter $A$ that plays the role of the width of the energy gap between the bands. Accompanying the shift of the control parameter $A$ from $-\infty$ to $+\infty$, an inversion of the two bands manifests itself through the exchange of two energy levels redistributing between the energy bands in opposite directions, see Fig.~\ref{ResumeGraf}(a). The quantum Hamiltonian describing the inversion of two bands is generalized to a model quantum Hamiltonian describing the inversion of a system of energy bands and then the global reorganization of bands associated with the band inversion phenomenon is explained in terms of redistribution of quantum energy levels.

The corresponding semi-quantum model is studied in Section~\ref{S_3} in order to reveal the topological origin of the redistribution phenomenon. In this model, the dynamical variables are split in two subsets of slow and fast variables. The slow variables corresponding to small transition frequencies are responsible for the internal structure of the bands and can be treated as classical variables. The fast variables are kept as quantum operators. The Hamiltonian is then represented as a matrix def\/ined on the classical slow variables. The eigenvalues of this semi-quantum matrix Hamiltonian determine the energy surfaces for the intra-band classical motion. The formation of degeneracy points of these eigenvalues, as schematically shown in Fig.~\ref{ResumeGraf}(b), is responsible for the modif\/ication of the topology of the eigenline bundles. The generic modif\/ications of the topological invariant, the Chern number, are explicitly calculated for both the model with and without the energy-ref\/lection symmetry. It is shown that the presence of the energy-ref\/lection symmetry imposes the simultaneous appearance of two degeneracy points with opposite delta-Chern contributions.

In Section~\ref{S_4}, the quantum multi-band problem is treated as a~completely classical one def\/ined on a phase space which is the product of two two-dimensional spheres. The classical Hamiltonian system is completely integrable because of the presence of the axial symmetry and can be characterized by the image of its energy-momentum map, see Fig.~\ref{ResumeGraf}(c). The energy-ref\/lection symmetry manifests itself through the $(E,J_z) \rightarrow (-E, -J_z)$ symmetry in the image of the energy-momentum map. A counterpart of the band inversion phenomenon consists in the qualitative modif\/ication of the system of critical values of the energy-momentum map occurring under the variation of the control parameter and is associated with the appearance of isolated critical values whose inverse image is a~pinched torus, implying the existence of Hamiltonian monodromy.

Returning back in Section~\ref{S_5} from the complete classical picture to the quantum problem, we are interested in the lattice of quantum states in the image of the energy-momentum map (see Fig.~\ref{ResumeGraf}(d)), which is def\/ined as the joint spectrum of the two commuting observables $(H=E,J_z)$. This joint spectrum for the transition region between $A\rightarrow -\infty$ and $A\rightarrow +\infty$ shows the presence of two elementary monodromy defects and indicates the splitting of the whole set of quantum states into bulk and edge states. The adjectives \emph{bulk} and \emph{edge} are to be understood in a spectral meaning: the bulk states correspond to energy levels that belong to the bulk of the energy bands while the edge states are associated with the energy levels redistributing between the bands during the band inversion process. In topological insulator theory, the word \emph{edge} has a spatial meaning. But the edge state plays also a~role of gap closing. In this sense, our nomenclature of edge state based on spectral theory is in accord with the spatial meaning since the edge states are responsible for the band rearrangement.

Section~\ref{S_6} presents the relation between the spectral f\/low for the multi-band quantum problem and the topological ef\/fects. These ef\/fects are discussed in the semi-quantum and completely classical analogs of the quantum problem. A generalization of the spectral f\/low notion to the band inversion phenomenon with an arbitrary number of bands is discussed.

Section~\ref{S_7} contains the conclusion.

\section{Quantum Hamiltonian and its symmetry} \label{S_Quantum}
We start in this section by considering a quantum system formed by ``slow'' and ``fast'' subsystems and assuming the existence of two quantum fast states such that further excitations of the fast subsystem is much more energy consuming than the excitations of the slow subsystem. In this situation we are allowed to write an ef\/fective Hamiltonian as a two-by-two Hermitian matrix with matrix elements being functions of quantum operators responsible for the internal structure of the two bands~\cite{BIZPhysRep}. We use the components of the angular momentum $L_\alpha$, $\alpha=x,y,z$ to describe the internal structure.

We assume that the ef\/fective quantum Hamiltonian respects the axial symmetry with diagonal action on the fast ($S_\alpha$, these variables are explicitly introduced in equation~(\ref{Hamiltonian_ref})) and slow ($L_\alpha$) variables (this means that both $S_\alpha$ and $L_\alpha$ transform as vectors) and contains $L_\alpha$ operators of the lowest degrees only. The Hamiltonian, consequently, takes the following simple form\footnote{The simplifying assumption of a diagonal action of the axial symmetry allows us to restrict of\/f-diagonal matrix elements to being linear in $L_\pm$. Note that a non-diagonal action leads to new interesting ef\/fects, in particular, to fractional monodromy for completely classical and fully quantum models \cite{Hansen, IwaiZhTheorChemAcc,Nekhoroshev}.}:
\begin{gather} \label{L+L-}
 H_{\rm quantum} = \left( \begin{matrix} A+\delta L_z +d L_z^2 & \bar{\gamma} L_- \\
 \gamma L_+ & -A -\delta L_z-d L_z^2 \end{matrix} \right),
\end{gather}
where $L_\pm= L_x\pm i L_y$, and where $A$ is a control parameter and $\delta$, $d$, $\gamma$ are phenomenological parameters among which $\delta$ and $d$ are real constants and $\gamma$ is a complex constant.

The space of quantum states on which this Hamiltonian acts is realized to be $\mathbb{C}^2\otimes \mathcal{H}(S^2)$, where $\mathbb{C}^2$ carries the fast (or vibrational-like) degrees of freedom and $\mathcal{H}(S^2)$ is the Hilbert space of square-integrable functions over the unit sphere $S^2$, carrying the slow (or rotational) degrees of freedom. Denoting a basis of $\mathbb{C}^2$ by $|V_+\rangle$ and $|V_-\rangle$ and a basis of $\mathcal{H}(S^2)$ by $|L,M_L\rangle$ with $|M_L|\leq L$ and $L=0,1,\dots$, we denote a~basis of $\mathbb{C}^2\otimes \mathcal{H}(S^2)$ by $|V_k;L,M_L\rangle$, $k=+,-$, where $|L,M_L\rangle$ are realized in terms of the spherical harmonics. Since $\mathbf{L}^2=L_x^2+L_y^2+L_z^2$ is an integral of motion, the label $L$ can be f\/ixed and the Hamiltonian~(\ref{L+L-}) is put in the form of an Hermitian matrix of size $2(2L+1)\times 2(2L+1)$ on the subspace spanned by $|V_k;L,M_L\rangle$, $k=+,-$, $|M_L|\leq L$.

Our main interest is in the analysis of qualitative modif\/ications of the band structure along the variation of the control parameter $A$. In other words, we are interested in the modif\/ication of the number of energy levels in the band which can take place according to the redistribution of the energy levels between bands under the variation of the control parameter $A$. In contrast to this modif\/ication, the variation of the internal structure of each band is of less importance, if the bands are isolated.

A quick observation is made on the Hamiltonian with a suf\/f\/iciently large control parameter. If $|A|$ is suf\/f\/iciently big, the of\/f-diagonal blocks of the Hamiltonian~(\ref{L+L-}) can be approximately neglected and then the Hamiltonian has two energy bands separated by an energy gap which becomes bigger than the width of each band for suf\/f\/iciently big $\left|A\right|$. In this case, the number of energy levels in both bands is the same, and equal to $2L+1$.

Now we set out to f\/ind eigenvalues of the Hamiltonian~(\ref{L+L-}) by the ef\/fective use of the ${\rm SO}(2)$ symmetry. The axial rotation about the $z$-axis gives rise to the unitary transformation on $\mathbb{C}^2\otimes \mathcal{H}(S^2)$ in the manner $U_t\Phi=e^{-it\frac12 \sigma_z}\Phi \circ R_{-t}$, where $\Phi$ is a two-component vector function on $S^2$ and where $\sigma_z$ is one of the Pauli matrices and $R_t$ denotes the axial rotation. We denote the inf\/initesimal generator of $U_t$ by $J_z$ or set $U_t=\exp(-itJ_z)$, then a~straightforward calculation provides
\begin{gather*}
J_z = \left( \begin{matrix} L_z+\frac12 & 0 \\ 0 & L_z-\frac12 \end{matrix} \right).
\end{gather*}The Hamiltonian~(\ref{L+L-}) is ${\rm SO}(2)$ invariant. In fact, one can easily verify that $[J_z, H_{\rm quantum} ]=0$. Hence, the eigenvalue problem for $H_{\rm quantum}$ reduces to subproblems on the eigenspaces of~$J_z$.

An alternative expression of the Hamiltonian~(\ref{L+L-}) is possible in terms of pseudo-spin components \cite{schwinger, pseudospin} with $S=1/2$ by associating with $|V_+\rangle$ and $|V_-\rangle$ states the ef\/fective projection of the pseudo-spin on the $z$-axis respectively equal to $+\frac12$ and $-\frac12$. The Hamiltonian can then be rewritten as
\begin{gather}
 H_{\rm quantum} = 2 S_z \otimes \big(A+\delta L_z+d L_z^2\big) + \gamma S_- \otimes L_+ + \bar{\gamma} S_+ \otimes L_-, \label{Hamiltonian_ref}
\end{gather}
where $S_{\pm}=S_x\pm iS_y$. In this representation, the pseudo-spin operators $S_\alpha$ act on the pseudo-spin component $|V_\pm\rangle$ and the angular momentum operators $L_\alpha$ act on the spherical harmonics $|L,M_L\rangle$. If we adopt this expression, it is easy to extend the Hamiltonian so as to act on the space of quantum states $\mathbb{C}^{2S+1}\otimes \mathcal{H}(S^2)$ and to keep the ${\rm SO}(2)$ symmetry, where the basis of~$\mathbb{C}^{2S+1}$ is denoted by $|V_k\rangle$, $|k|\leq S$, and the basis of $\mathbb{C}^{2S+1}\otimes \mathcal{H}(S^2)$ by $|V_k;L,M_L\rangle$, and where the pseudo-spin operators $S_{\alpha}$ are to be represented in the form of $(2S+1)\times (2S+1)$ matrices. The ${\rm SO}(2)$ symmetry operator is given by the unitary operator $U_t=e^{-itS_z}\otimes e^{-itL_z}$, which acts on $S_\pm$ and $L_\pm$ in the manner
\begin{gather*}
e^{-itS_z}S_\pm e^{itS_z} = e^{\mp it}S_\pm, \qquad e^{-itL_z}L_\pm e^{itL_z} = e^{\mp it}L_\pm,
\end{gather*}and leaves $S_z$ and $L_z$ invariant, so that one has $U_t H_{\mathrm{quantum}} U_t^{-1} = H_{\mathrm{quantum}}$. The inf\/initesimal generator of $U_t = e^{-itS_z}\otimes e^{-itL_z}$ is shown to be put in the form $J_z=S_z\otimes I + I\otimes L_z$, where~$I$ denotes the identity of the respective factor spaces of $\mathbb{C}^{2S+1}\otimes {\cal{H}}(S^2)$. However, we denote the operator $J_z$ by $J_z=L_z+S_z$ for notational simplicity. Then, one immediately verif\/ies that $[J_z, H_{\mathrm{quantum}}]=0$.

\looseness=-1 Postponing the study of the case of a generic $S$, we here deal f\/irst with the case of $S=\frac12$. The ${\rm SO}(2)$ symmetry allows us to represent the Hamiltonian in a block-diagonalized form with two one-dimensional blocks associated with invariant subspaces of both $H_{\rm quantum}$ and $J_z$ given by
\begin{gather*}
|V_-;L,M_L=-L\rangle \qquad {\rm and} \qquad |V_+;L,M_L=L\rangle, \end{gather*}respectively, and $2L$ two-dimensional blocks associated with two-dimensional invariant subspaces (again of both $H_{\rm quantum}$ and $J_z$ operators) spanned by
\begin{gather*}
|V_-;L,M_L\rangle, \qquad |V_+; L,M_L-1\rangle, \qquad M_L=L, L-1, \ldots, -L+1.
\end{gather*}Thus, our eigenvalue problem amounts to f\/inding eigenvalues of Hermitian matrices representing $H_{\rm quantum}$ on these invariant subspaces, and hence the explicit expressions of the eigenstates are easily obtained for arbitrary choice of values of the phenomenological parameters $\delta$, $d$, $\gamma$ and the control parameter~$A$.

The eigenvalues for the two one-dimensional blocks are
\begin{gather}
H_{\rm quantum} |V_-; L,M_L =-L\rangle = \big({-}A+\delta L - d L^2\big) |V_-; L,M_L=-L\rangle, \label{new1a}\\
H_{\rm quantum} |V_+; L, M_L =L\rangle =\big(A+\delta L + d L^2\big) |V_+; L,M_L=L\rangle.\label{new1b}
\end{gather}
It is to be noted that these two eigenvalues are linear in $A$, which consequently implies that under variation of the control parameter~$A$ from big negative to big positive values, these two eigenvalues go from one band to another and participate in the redistribution of the energy levels between bands.

The matrix expression of the Hamiltonian in each two-dimensional invariant subspace,
\begin{gather*}
 {\rm span}\big\{ |V_-; L,M_L\rangle, |V_+; L,M_L-1\rangle \big\},
\end{gather*}is given by
\begin{gather}
 \label{AddRef2block}
 \left( \begin{matrix} -\big(A + \delta M_L + d M_L^2\big) & \gamma \sqrt{L(L+1)-M_L(M_L-1)}\vspace{1mm}\\
 \bar{\gamma} \sqrt{L(L+1)-M_L(M_L-1)} & A + \delta (M_L-1) + d (M_L-1)^2
 \end{matrix} \right).
\end{gather}
The eigenvalues of the present block Hermitian matrix are
\begin{gather}
 \lambda^{\pm}(L,M_L) = -\frac{\delta+(2M_L-1)d}{2} \nonumber \\
{} \pm \sqrt{ \big[ A+\delta \big(M_L-\tfrac{1}{2}\big) + d\big(M_L^2-M_L+\tfrac{1}{2} \big) \big]^2 + |\gamma |^2 ( L(L+1)-M_L(M_L-1) ) }.\label{2dblocks}
\end{gather}
The eigenvalues $\lambda^+(L,M_L)$ and $\lambda^-(L,M_L)$ of the two-dimensional blocks (\ref{AddRef2block}) given by equation~(\ref{2dblocks}) do not cross each other, accompanying the variation of the control parameter $A$, if~$M_L$ is f\/ixed with $M_L=L, L-1, \ldots, -L+1$. This implies that the eigenvalues $\lambda^+(L,M_L)$ and $\lambda^-(L,M_L)$ with~$M_L$ ranging from $-L+1$ to $+L$ in (\ref{AddRef2block}) form respective bands, upper and lower, without referring to the eigenvalues given in equations~(\ref{new1a}) and (\ref{new1b}), which are responsible for the redistribution of the energy levels. However, for suf\/f\/iciently large values of $|A|$, there are two energy bands to one of which each eigenvalue from equations~(\ref{new1a}) and (\ref{new1b}) belongs. In view of this, we are allowed to introduce a label $b$ to assign each of the bands. This label takes integer values starting with $b=0$ and increasing along with the increasing order in energy. The Hamiltonian we are dealing with has two bands, for which the label $b$ takes two values: $b=0$ and $b=1$ for the lower and the upper energy bands, respectively.

\begin{figure}[t]\centering
 \includegraphics[scale=0.78]{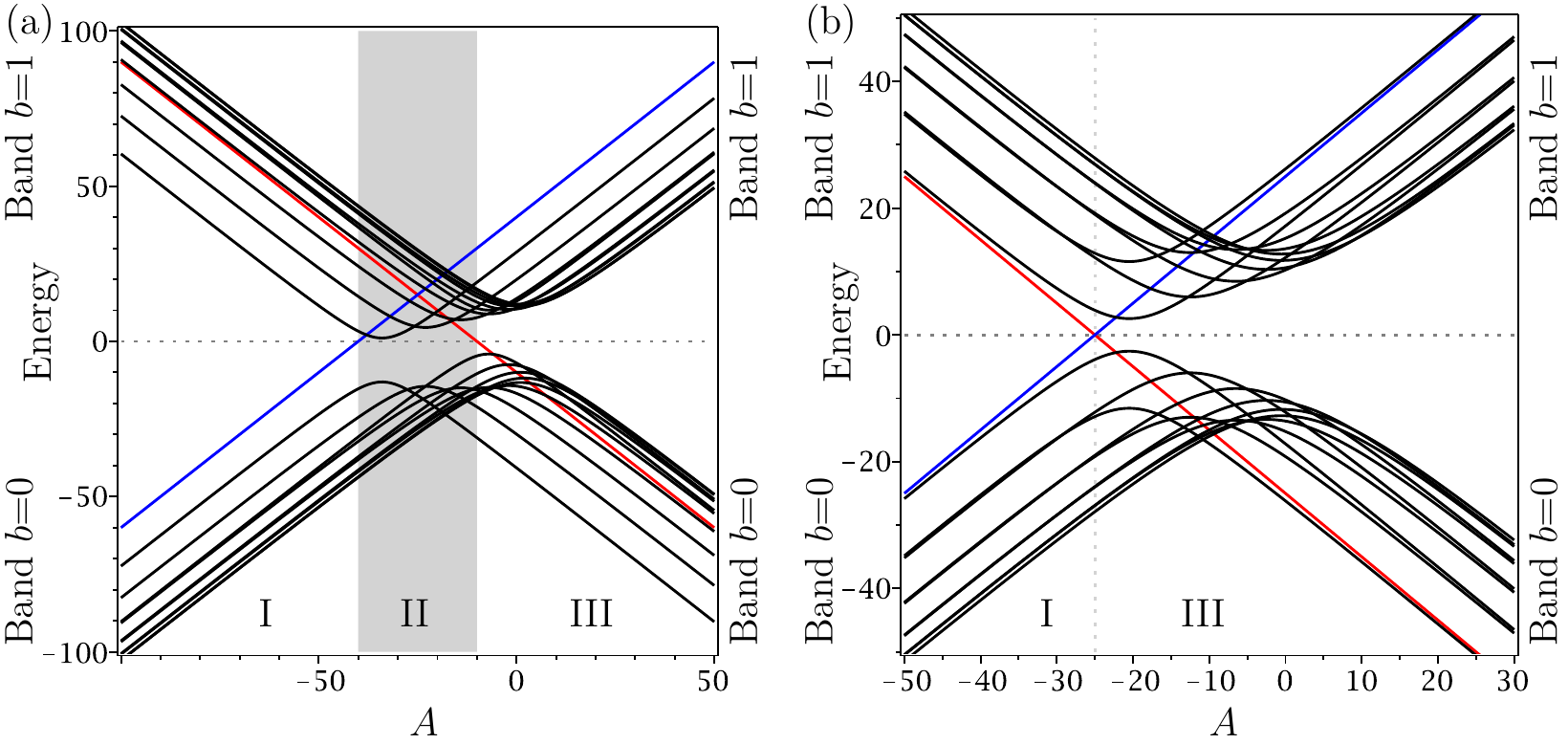}
\caption{Evolution of the pattern of quantum energy levels of the Hamiltonian (\ref{L+L-}) under variation of control parameter $A$. The blue and red lines correspond to the two one-dimensional blocks (edge states). The black lines are associated with the solutions of the $2\times2$ blocks, see equation~(\ref{2dblocks}), and can be associated to the bulk states of each band. Symbols I, II, and III indicates intervals of $A$-values corresponding to three dif\/ferent iso-Chern domains. Figures are done for the following choice of phenomenological parameters of the Hamiltonian~(\ref{L+L-}): (a) $L=5$, $\gamma = 1+2i$, $d=1$, $\delta=3$. (b) $L=5$, $\gamma = 1+2i$, $d=1$, $\delta=0$. \label{Z2axialDelta} }
\end{figure}

An example of the evolution of the energy level pattern along with the variation of the control parameter $A$ is shown in Fig.~\ref{Z2axialDelta}(a). We note the existence of two bands: for big $\left|A\right|$, the eigenvalues $\lambda^{-}(L,M_L)$ belong to the lower band, while the eigenvalues $\lambda^{+}(L,M_L)$ belong to the upper band. The average $\langle S_z \rangle$ of the projection of the pseudo-spin operator on the $z$ axis can be used to characterize these bands in the limit of big $|A|$. The lower and upper bands are respectively associated with $\langle S_z \rangle=-1/2$ and $\langle S_z \rangle=+1/2$ for big positive $A$ values (domain~III) and the correspondence is reversed in the limit $A\rightarrow -\infty$ (domain~I). For intermediate values of $A$ (domain~II), the average value of $S_z$ varies, which means that a band rearrangement is in progress under the variation of the control parameter $A$, and hence the average value of $S_z$ cannot be used to label bands. However, the energy levels $\lambda^+(L,M_L)$ and $\lambda^-(L,M_L)$ belong to the upper and lower bands, respectively, independently of the variation of the control parameter. Such states that always belong to the same band are called bulk states.

The two levels represented by red and blue lines in Fig.~\ref{Z2axialDelta}(a) change bands along the variation of $A$, they are the so-called edge states. The redistribution between bands of these two energy levels occurs at two dif\/ferent $A$-values with the dif\/ference between these values depending on the parameter $\delta$. The position of these levels with respect to $E=0$ can be used to assign these two levels to one or to another band. If $\delta$ is positive as in Fig.~\ref{Z2axialDelta}(a) the upward blue quantum energy level crosses the horizontal axis $E=0$ before the downward red level, but the order in the crossing is reversed for negative $\delta$. In the part of Fig.~\ref{Z2axialDelta}(a) with a white background (domains~I and~III), one level is in the lower band and the other level is in the upper band, while the two blue and red levels belong to the upper band in the grey area (domain II). As a consequence, the number of levels in each band changes with $A$: for $\delta>0$, it is $2L+1$ in the domain~I and~III, while it is $2L+2$ in the upper band and $2L$ in the lower band in the domain~II. In case of a~negative value of $\delta$, the blue and the red lines cross each other in the region of $E < 0$. Then, the number of levels of the upper band is $2L$ and that of levels of the lower band is $2L+2$ in the domain~II, while in the domains~I and III the number of levels of the upper and the lower bands are the same and equal to $2L+1$. An alternative graphical representation of the redistribution phenomenon in terms of the evolution of the joint spectrum of two commuting observables, $(E,J_z)$, will be discussed in Section~\ref{S_5}.

\subsection{Additional f\/inite symmetry and pseudo-symmetry}
If we impose $\delta=0$, the eigenvalue pattern of the Hamiltonian~(\ref{L+L-}) becomes more symmetric, see Fig.~\ref{Z2axialDelta}(b). This can be seen from the explicit expressions for eigenvalues which for $\delta=0$ become
\begin{gather*}
E(A, \delta=0, L, M_L=-L) = -A - d L^2, \\
E(A, \delta=0, L, M_L=L) = A + d L^2, \\
E^\pm(A, \delta=0, L, M) = \big({-}M_L+\tfrac 12\big)d \\
\hphantom{E^\pm(A, \delta=0, L, M) =}{}\pm \sqrt{\big[A+ d\big(M_L^2-M_L +\tfrac12\big)\big]^2 + |\gamma|^2 ( L(L+1)-M_L(M_L-1) ) }.
\end{gather*}
One can easily verify that
\begin{gather*}
E^\pm(A;\delta=0, L,-M_L+1) = -E^\mp(A; \delta=0, L,M_L).
\end{gather*}This means that the energy level pattern satisf\/ies the energy-ref\/lection symmetry, i.e., if \smash{$E(\delta=0)$} is an eigenvalue of the Hamiltonian~(\ref{L+L-}), then $-E$ is also an eigenvalue. The right border of domain~I and the left border of domain~III coincide for $\delta=0$ and the domain~II is empty.

The energy-ref\/lection symmetry can be explained without any reference to the explicit form of the eigenvalues by applying the following antiunitary transformation to the Hamiltonian $H_{\rm quantum}$ with $\delta=0$. First we def\/ine the action of complex conjugation~$K$ on Hamiltonian~(\ref{L+L-}) as consisting in the transformation $L_\alpha \rightarrow - L_\alpha$ and in the complex conjugation of all coef\/f\/icients. This means that the dif\/ferent terms of (\ref{L+L-}) transform according to
\begin{gather} \label{L variable transform}
\gamma L_+ \rightarrow - \bar{\gamma} L_- , \qquad \bar{\gamma} L_- \rightarrow - \gamma L_+, \qquad L_z \rightarrow - L_z, \qquad L_z^2 \rightarrow L_z^2.
\end{gather}
Adding the unitary transformation with matrix $\sigma_1=\sigma_1^{-1} =\left(\begin{smallmatrix} 0&1\\1&0\end{smallmatrix}\right)$, we easily verify
 for Hamiltonian (\ref{L+L-}) that
\begin{gather}
\label{TransfGen}
 (\sigma_1 K)H_{\rm quantum} (\sigma_1 K)^{-1} =-H_{\rm quantum}, \qquad {\rm with} \quad \delta =0.
\end{gather}
This implies that if $\Phi$ is an eigenstate associated with an eigenvalue $E$ then $\sigma_1 K\Phi$ is an eigenstate associated with the eigenvalue $-E$. The action of the same transformation on the basis functions yields the mapping between invariant subspaces for $H_{\rm quantum}$,
\begin{gather*}
 {\rm span}\big\{|V_-; L, M_L\rangle, |V_+; L,M_L-1\rangle\big\} \rightarrow {\rm span}\big\{|V_+; L, -M_L+1\rangle, |V_-; L,-M_L\rangle\big\}.
\end{gather*}This shows that the antiunitary transformation $\sigma_1 K$ gives rise to the substitution $M_L \mapsto -M_L+1$ in the indices of the basis vectors. Using the fact that the Hamiltonian with $\delta=0$ changes the sign under the same transformation, we conclude that in the case of $\delta=0$, the energy level pattern for $H_{\rm quantum}(\delta=0)$ respects the energy-ref\/lection symmetry which is a characteristic symmetry transformation for supersymmetric quantum mechanics~\cite{cooper}. Fig.~\ref{Z2axialDelta}(b) illustrates this symmetry on a concrete example.

It should be noted that the energy-ref\/lection symmetry can sometimes be realized in an other way, say, by using a unitary transformation. A simple molecular example with such a property is the maximally asymmetric rigid rotor described by the Hamiltonian $H_\mathrm{asym}=B(L_x^2-L_y^2)$. The invariance group for a generic asymmetric rotor is $D_{2h}$. The Hamiltonian $H_\mathrm{asym}$ in addition transforms according to a one-dimensional not totally symmetric representation of the $D_{4h}$ group. In particular $H_\mathrm{asym}$ changes sign under the rotation by $\pi/2$ around the $z$ axis. The same unitary transformation interchanges eigenfunctions of $H_\mathrm{asym}$. This implies the energy-ref\/lection symmetry for $H_\mathrm{asym}$. However, the main topic of the present analysis are ef\/fective Hamiltonians which change the sign under an antiunitary operator $U_C K$ represented as a~product of a unitary operator $U_C$ and the complex conjugation $K$, like $(U_CK){H} (U_C K)^{-1}=-H$.

An even more symmetric energy level pattern is obtained for Hamiltonian~(\ref{L+L-}) if we impose $d=0$ in addition to $\delta=0$. The energy level pattern in such a case reveals an additional parametric symmetry. It is invariant under the $A\rightarrow -A$ ref\/lection. We used precisely this case to illustrate the rearrangement of energy levels for a quantum two-band problem in Fig.~\ref{ResumeGraf}(a).

\subsection{Band inversion for an arbitrary number of bands\label{S22}}
We proceed to the case of $S$ being an arbitrary positive integer or half-integer with a main interest in the inversion of the system of $2S+1$ bands for the Hamiltonian~(\ref{Hamiltonian_ref}) under the variation of the control parameter $A$ from very big negative to very big positive $A$ values. As is easily seen, the Hamiltonian in the two limits ($A \rightarrow \pm \infty$) does not depend on the details of the Hamiltonian, since the contribution of the $AS_z$ term is dominant in the limit of big $|A|$. This allows us to characterize the energy bands in the big $|A|$ limit by an average value of $S_z$ and to see that under the change of the sign of $A$ the order of the energy bands is inversed. At the same time we can label bands by consecutive integers $b=0,1,2,\ldots,2S$ in a similar manner to that done earlier for the two-band model. This band label does not change under variation of the control parameter $A$.

In order to solve the eigenvalue problem for $H_{\rm quantum}$ with $N=2S+1$ bands in the presence of the axial symmetry and under the assumption that $S\ll L$, we can split the problem into subproblems on respective eigenspaces for $J_z=L_z+S_z$. This is because each eigenspace of~$J_z$ is an invariant subspace of $H_{\rm quantum}$ on account of $[J_z,H_{\rm quantum}]=0$.

\begin{figure}[t]\centering
 \includegraphics[scale=0.80]{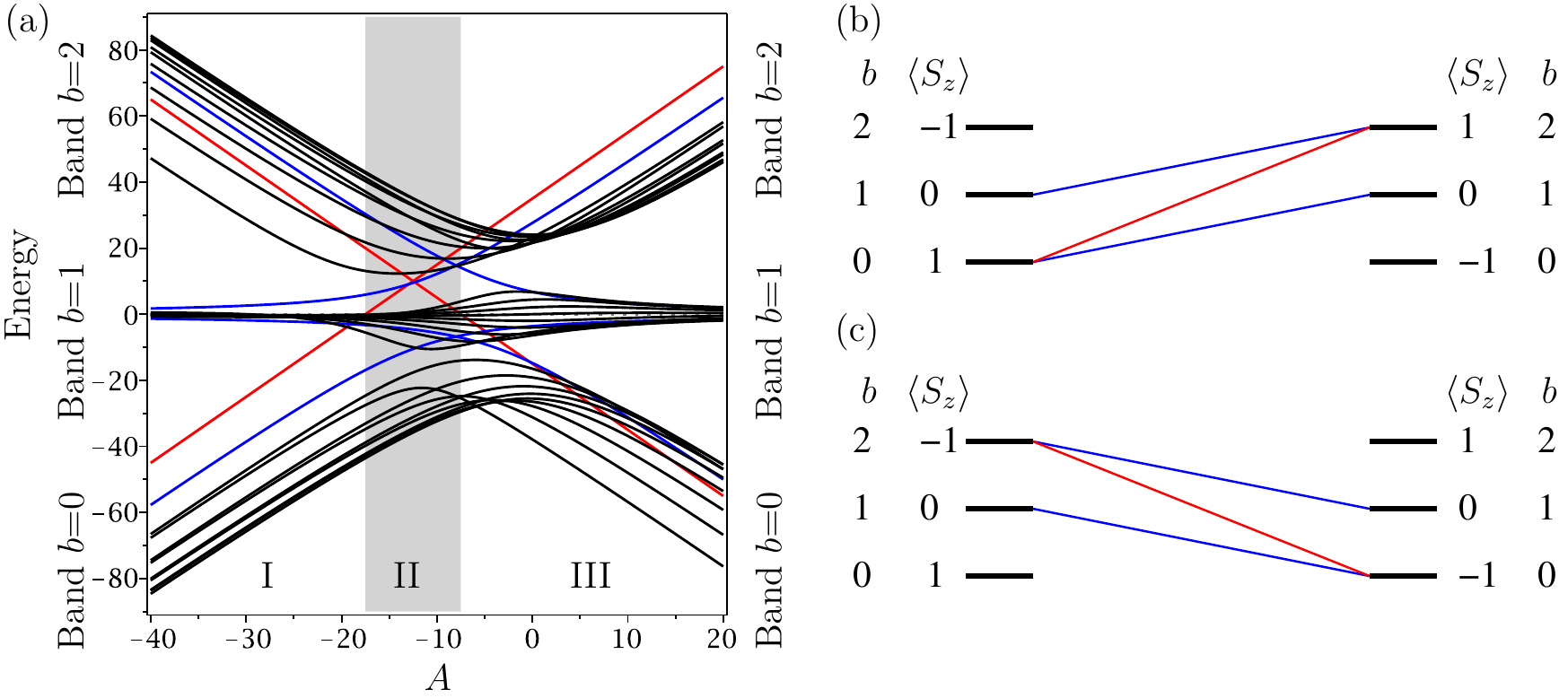}
\caption{Quantum energy level pattern for the $S=1$ problem with $L=5$, $g=1+2i$, $d=1/2$, $\delta=1$. The edge states in red and blue respectively belong to the one-dimensional invariant subspaces $J_z=\pm (L+S)$ and the two-dimensional invariant subspaces $J_z=\pm (L+S-1)$. (a) General view of the quantum energy level pattern. Symbols~I, II, and~III indicates intervals of $A$-values corresponding to three dif\/ferent iso-Chern domains. (b) Correlation diagram showing the redistribution of the energy levels that change band in the upward direction between the $A\rightarrow -\infty$ and the $A\rightarrow \infty$ limits. In each limit, the bands can be labeled by increasing energy with an integer $b=0,1,2$, or by the average value of $\langle S_z \rangle$. Only the levels which change bands under control parameter $A$ variation are shown. (c) Same as (b) but for the energy levels that change band in the downward direction between the $A\rightarrow -\infty$ and the $A\rightarrow \infty$ limits.}\label{FigG34}
\end{figure}

For given $L$ and $S$, the eigenvalues of $J_z$ are integers or half-integers ranging from $-L-S$ to $L+S$. In what follows, we use the same symbol $J_z$ to denote its eigenvalues for notational simplicity. The eigenspace associated with an eigenvalue $J_z$ ($|J_z|\leq L+S$) is spanned by states $|V_k;L,M_L\rangle$ with $k+M_L=J_z$. Explicitly speaking, the eigenspaces associated with $J_z=\pm (L+S)$ and $J_z=\pm (L+S-1)$ are of dimension one and two, respectively, and so on. The eigenspace with $J_z=\pm (L-S+1)$ is of dimension $2S$. For $-L+S \leq J_z\leq L-S$, the eigenspaces are of maximal dimension $2S+1$.

We assign invariant subspaces of the Hamiltonian by using the eigenvalues of $J_z$. Under variation of the control parameter~$A$, the eigenvalues obtained from the subproblem on the invariant subspace with the $J_z$ eigenvalue f\/ixed do not cross one another in general, since the Hermitian matrices determined on the respective invariant subspaces have one continuous parameter $A$ and at most one discrete parameter~$M_L$ with~$L$ and~$S$ f\/ixed, and since the codimension of degeneracy in eigenvalues is three for any Hermitian matrix of size greater than or equal to two~\cite{ArnoldCodim, Neumann}. In particular, energy eigenvalues from the blocks with $-L+S \leq J_z\leq L-S$ do not cross one another under variation of the control parameter and the number of energy levels in each block is maximal or~$2S+1$.

Hence, the totality of the energy levels associated with bulk states form bands with labels $b=0,1,\ldots,2S$ in the increasing order in energy. The bulk states have nothing to do with band rearrangement and the labels assigned to respective bands are constant in~$A$. In contrast with this, eigenvalues coming from dif\/ferent invariant subspaces of the Hamiltonian may cross one another and energy levels from blocks of smaller size with $|J_z|>|L-S|$ are responsible for band rearrangement. In the limit of suf\/f\/iciently big $|A|$, we may extend the labeling of bands so as to include the energy levels for edge states. In this situation, for an arbitrary $S\ll L$, the redistribution of energy levels takes place in the manner to be stated below.

The energy levels belonging to the one-dimensional invariant subspaces with $J_z=\pm(L+S)$ should go from $b=S \mp S$ bands to $b=S \pm S$ bands, i.e., from the lowest in energy band to the highest in energy band and vice versa. These levels change the band label by $\pm 2S$. Energy levels belonging to the two-dimensional invariant subspaces with $J_z=\pm (L+S-1)$ should go from $b=S \mp(S-c)$ bands with $c=0,1$ to bands with $b=S \pm (S+c-1)$, i.e., they change the band label by~$\pm(2S-1)$. More generally, energy levels belonging to $m$-dimensional invariant subspaces, $m=1,2,\ldots, 2S$, associated with $J_z=\pm (L+S-m + 1)$ should go from $b=S \mp(S-c)$ bands with $c=0,1, \dots , m-1$ to bands with $b=S \pm (S+c-m+1)$, i.e., they change the band label by $2S-m+1$. Thus, all energy eigenvalues belonging to the invariant subspace of non-maximal dimension for $J_z$ should change energy bands under the variation of the control parameter from big negative to big positive $A$ values. We call this phenomenon the inversion of the whole set of bands because the average value~$\langle S_z \rangle$ changes the sign under variation of the sign of~$A$ in the limit of big~$|A|$ for the band with label~$b$.

\begin{figure}[t]\centering
 \includegraphics[scale=0.8]{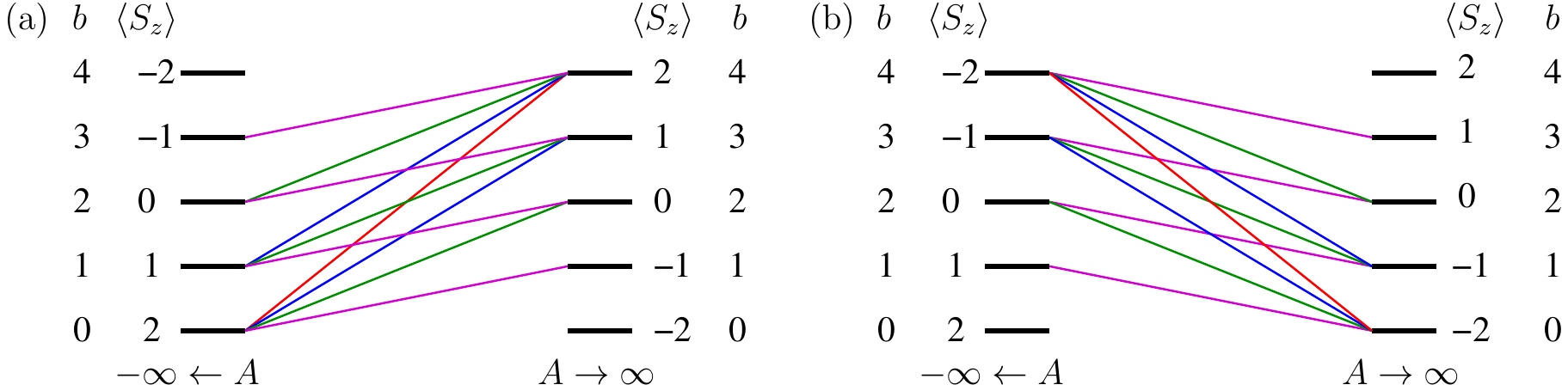}
\caption{Correlation diagram showing the redistribution of the energy levels between bands associated with an inversion of the system of bands. Example of the system of f\/ive bands for the case of an ef\/fective spin $S=2$. The bands symbolized by a horizontal thick line are labelled by the average value of $\langle S_z \rangle$ and by a quantum number $b=0,1,2,3,4$ attributed consecutively to bands with increasing energy. Only the energy levels which change bands under the variation of the control parameter $A$ are shown. To make the f\/igure more easy to read, the levels belonging to the invariant subspaces $J_z=-L-S_z$ and $J_z=L+S_z$ ($S_z=S, S-1, \dots , -S+1$) are shown in separate subf\/igures (a) and (b). Levels belonging to invariant subspaces of the same dimension are shown by the same color. Red color: one-dimensional invariant subspaces, $J_z=\pm (L+S)$. Blue color: two-dimensional invariant subspaces, $J_z=\pm (L+S-1)$. Green color: three-dimensional invariant subspaces, $J_z=\pm (L+S-2)$. Magenta color: four-dimensional invariant subspaces, $J_z=\pm (L+S-3)$. \label{RearrangS} }
\end{figure}

Fig.~\ref{FigG34} illustrates the rearrangement of the energy levels between bands for the example of the Hamiltonian~(\ref{Hamiltonian_ref}) with $S=1$. In particular, Fig.~\ref{FigG34}(b,c) illustrate schematically the global rearrangement of energy levels by showing the correlation diagram for the case $S=1$ with three energy bands exhibiting global inversion of the band system. Fig.~\ref{RearrangS} presents in a similar way the inversion of the band system for $S=2$ and f\/ive energy bands.

\section{Semi-quantum model and Chern numbers} \label{S_3}
The redistribution of energy levels for the Hamiltonian~(\ref{L+L-}) against the control parameter has a counterpart in the semi-quantum limit~\cite{PhysLettMono}. In the limiting procedure, the rotational variables are viewed as slow variables and then treated as classical ones and the fast vibrational-like variables remain to be quantum ones, but only a small (two in the simplest case) number of quantum states are taken into account. The matrix Hamiltonian~(\ref{L+L-}) then becomes to be def\/ined on the classical phase space for slow variables. In order to stress the dif\/ference between the quantum and the semi-quantum versions, we denote below the slow variables by $x_1$, $x_2$, $x_3$ instead of $L_x$, $L_y$, $L_z$, hence the semi-quantum Hamiltonian is put in the form
\begin{gather} \label{L+L-SemiQuant}
H_{\text{\rm semi-quantum}} = \left( \begin{matrix} A+\delta x_3 +d x_3^2 & \bar{\gamma} (x_1-i x_2) \\
 \gamma (x_1+i x_2)& -A -\delta x_3-d x_3^2 \end{matrix} \right),
\end{gather}
with the renormalized restriction on~$x_k$ variables $x_1^2+x_2^2+x_3^2=1$. The classical phase space is thus a two-dimensional unit sphere~$S^2$ in~$\mathbb{R}^3$, the space of~$x_k$ variables. The two eigenvalues of $H_{\text{\rm semi-quantum}}$ (\ref{L+L-SemiQuant}) are functions def\/ined on~$S^2$, depending on the control parameter~$A$ and on the phenomenological parameters. If the eigenvalues are not degenerate, each of two associated eigenspaces is assigned to every point of~$S^2$ to form a complex line bundle, called an eigenline bundle, over~$S^2$.

We now wish to relate the qualitative modif\/ications occurring in the quantum energy level pattern for $H_{\rm quantum}$ under the variation of the control parameter $A$ to the qualitative modif\/ications observed for a f\/iber bundle associated with the semi-quantum model under the same variation of the control parameter $A$. If the eigenvalues remain non-degenerate on the whole base space against $A$, the eigenline bundles remain topologically unchanged, so that no topological phase transition of the band structure is expected. Thus, we need to f\/ind the degeneracy points of eigenvalues which can appear during the variation of the control parameter $A$ and to characterize the modif\/ications occurring with eigenline bundles when the control parameter goes through points associated with degeneracy in eigenvalues.

Let us start by looking at the symmetry of the semi-quantum Hamiltonian~(\ref{L+L-SemiQuant}). The semi-quantum Hamiltonian possesses the same axial symmetry as its quantum analog, which is put
in the form
\begin{gather}
\label{sq SO(2) symmetry}
 e^{-it\sigma_z/2}H_{\text{\rm s-q}}(x_1+ix_2,x_1-ix_2,x_3)e^{it\sigma_z/2} = H_{\text{\rm s-q}}\big(e^{it}(x_1+ix_2), e^{-it}(x_1-ix_2),x_3\big),
\end{gather}
 where the subscript $\text{\rm s-q}$ is the abbreviation of semi-quantum. Under the action of the axial symmetry, the base space $x_1^2+x_2^2+x_3^2=1$ is stratif\/ied into orbits. There are two isolated orbits, at the north and south poles of the sphere, which are invariant under axial symmetry. All other orbits are one-dimensional circles. If a point is a degeneracy point of the eigenvalues, all points of its orbit should necessarily be degeneracy points, as is seen from~(\ref{sq SO(2) symmetry}). For this reason, a~circle of constant latitude could be a set of degeneracy points. However, this rarely happens on account of the codimension condition. Appearance of a~degeneracy at a circular orbit for a~one-parameter family of~${\rm SO}(2)$ invariant Hamiltonians is not generic. We are here reminded that for a Hermitian matrix, the codimension of degeneracy is three~\cite{ArnoldCodim, Neumann} and, consequently, for a~one-parameter family of semi-quantum Hamiltonians def\/ined on a two-dimensional base space, degeneracy points generically exist and are isolated. It then follows that isolated degeneracy points under the presence of the axial symmetry can appear only at the north pole, at the south pole, or at both of them.

The two eigenvalues of the semi-quantum Hamiltonian~(\ref{L+L-SemiQuant}) are
\begin{gather} \label{eigenSemiQ}
 \pm \sqrt{\big(A+\delta x_3+ d x_3^2\big)^2 + |\gamma|^2 \big(x_1^2+x_2^2\big) }.
\end{gather}
It is clear from this expression that the two eigenvalues are symmetric with respect to the $E=0$ axis for Hamiltonians with arbitrary~$\delta$, i.e., the traceless semi-quantum two-level Hamiltonian always satisf\/ies the energy-ref\/lection symmetry even though its quantum analog does not respect this symmetry. Taking into account the fact that the points of the base space for semi-quantum Hamiltonian are real, we can express the energy-ref\/lection symmetry for the semi-quantum Hamiltonian as a result of the transformation
\begin{gather}\label{energy-reflection symmetry, SQ}
i \sigma_2 \overline{H_{\text{\rm semi-quantum}}} (-i \sigma_2) = -H_{\text{\rm semi-quantum}},
\qquad \text{with} \quad i\sigma_2 = \left(\begin{matrix} 0&1\\-1&0\end{matrix} \right).
\end{gather}

From equation~(\ref{eigenSemiQ}) it follows that the degeneracy in eigenvalues of the semi-quantum Hamiltonian~(\ref{L+L-SemiQuant}) occurs for $\gamma\neq 0$ if and only if
\begin{gather}
 x_1=x_2=0, \qquad A+\delta x_3 + d x_3^2 =0. \label{Cond3}
\end{gather}
If $x_1=x_2=0$ then $x_3^2=1$, so that the second of the above equations (\ref{Cond3}) becomes for $x_3=+1$: $A+ \delta +d =0$ and for $x_3=-1$: $A-\delta+d=0$. This means that the degeneracy occurs, when $A=-d-\delta$, at the north pole of the unit sphere and when $A=-d+\delta$ at the south pole of the unit sphere. If the phenomenological parameter of the Hamiltonian $\delta \rightarrow 0$ tends to zero, both degeneracy points occur at the same value of the control parameter $A$ but at dif\/ferent poles of the unit sphere.

In order to simplify the analysis of the evolution of the eigenline bundles under the variation of the control parameter $A$, we consider independently two cases. In the f\/irst case, we study the Hamiltonian $H_{\text{\rm semi-quantum}}$ with $d=0$, $\delta=1$ or $\delta=-1$. Two degeneracy points appear in this case at $A_{\rm deg}=\pm1$ at dif\/ferent poles of the sphere. In the second case, we look at the Hamiltonian $H_{\text{\rm semi-quantum}}$ with $d=1$ and $\delta=0$. In that case, two degeneracy points appear at the same value of the control parameter $A_{\rm deg}=-1$ at dif\/ferent poles of the sphere.

For both cases we calculate for each eigenline bundle depending on the control parameter $A$ the modif\/ication of the topological invariant, the Chern number, associated with crossing the point $A_{\rm deg}$ associated with the eigenvalue degeneracy. The method of calculation is explained in details in~\cite{AnnPhysIZh}. We just summarize here the most important steps.

\subsection[Chern numbers for $H_{\text{\rm semi-quantum}}$ with $d=0$]{Chern numbers for $\boldsymbol{H_{\text{\rm\bf semi-quantum}}}$ with $\boldsymbol{d=0}$}

\subsubsection[Chern numbers for $H_{\text{\rm semi-quantum}}$ with $d=0$, $\delta=1$]{Chern numbers for $\boldsymbol{H_{\text{\rm\bf semi-quantum}}}$ with $\boldsymbol{d=0}$, $\boldsymbol{\delta=1}$}

Let us denote the eigenvalues of the Hamiltonian~$H_{\text{\rm semi-quantum}}$ with $d=0$, $\delta=1$ by~$\mu^\pm$:
\begin{gather*}
\mu^\pm = \pm \sqrt{(A+x_3)^2 + |\gamma|^2\big(x_1^2+x_2^2\big) } .
\end{gather*}The degeneracies occur at ${\mathbf{e}}_3$, the north pole, when $A=-1$ and at ${\mathbf{- e}}_3$, the south pole, when $A=1$. From the eigenvalue equation for $H_{\text{\rm semi-quantum}}$, we obtain in two ways the eigenvectors associated with~$\mu^+$:
\begin{alignat*}{3}
 & |v_{\rm up}^+\rangle = \frac{1}{\tilde{N}_{\rm up}^+}
 \left( \begin{matrix}\bar{\gamma}(x_1-ix_2) \\
 \mu_+ - (A+x_3) \end{matrix}\right), \qquad && \tilde{N}_{\rm up}^+ = \sqrt{ 2\mu_+(\mu_+ - (A+x_3)) }, & \\
& |v_{\rm down}^+\rangle = \frac{1}{\tilde{N}_{\rm down}^+}
 \left( \begin{matrix}\mu_+ + A+x_3 \\ \gamma(x_1+ix_2)\end{matrix}\right),
 \qquad && \tilde{N}_{\rm down}^+ = \sqrt{ 2\mu_+(\mu_+ + A+x_3) }. &
\end{alignat*}
The exceptional points for $|v_{\rm up}^+\rangle$ and $|v_{\rm down}^+\rangle$ (i.e., points where eigenvectors are not def\/ined) are determined by the equations $\{x_1=x_2=0, \, \mu^+ - (A+x_3)=0\}$ and $\{x_1=x_2=0,\, \mu^+ + A+x_3 =0\}$, respectively. The exceptional points are listed as follows:
\begin{gather} \label{TableExP1}
\begin{array}{c|c|c|c|c|c}
 &A <-1 & A=-1 & -1<A<1 & A=1 & A>1 \\ \hline
{\rm except.\ pts. \ for} \ |v_{\rm up}^+\rangle & {\rm no} &
\left(\begin{matrix} \mathbf{e}_3\\ {\rm deg.\ pt.} \end{matrix} \right) &
\mathbf{e}_3 & \mathbf{e}_3 & \pm\mathbf{e}_3 \\ \hline
{\rm except.\ pts. \ for} \ |v_{\rm down}^+\rangle & \pm\mathbf{e}_3 &
-\mathbf{e}_3 &-\mathbf{e}_3 &
\left(\begin{matrix} -\mathbf{e}_3\\ {\rm deg.\ pt.} \end{matrix} \right) &
{\rm no} \\
\end{array}
\end{gather}
If an eigenvector is globally def\/ined on the sphere, the eigenline bundle is trivial. If the bundle is not trivial, eigenvectors are def\/ined locally and the locally def\/ined eigenvectors are related together by a structure group (or a gauge group). The existence of exceptional points usually means that the locally def\/ined eigenvector can not be extended continuously on the whole two-sphere.

Table~(\ref{TableExP1}) shows that for $A < -1$ the eigenvector $|v_{\rm up}^+\rangle$ is globally def\/ined and for $A > 1$ the eigenvector $|v_{\rm down}^+\rangle$ is globally def\/ined, so that the eigenline bundle associated with $\mu^+$ is trivial for $A < -1$ and for $A > 1$ and hence the Chern number of the eigenline bundle in question is zero. However, the eigenline bundle associated with $\mu^+$ is non-trivial for $-1 < A < 1$. The eigenvectors $|v_{\rm up}^+\rangle$ and $|v_{\rm down}^+\rangle$ are related on the intersection of their domains by
\begin{gather*}
|v_{\rm up}^+\rangle = \eta |v_{\rm down}^+\rangle, \qquad \eta = \frac{\bar{\gamma}(x_1-ix_2)}{|\gamma| \sqrt{x_1^2+x_2^2}}.
\end{gather*}

To calculate the Chern numbers of the eigenline bundle associated with $\mu^+$ for $-1<A<1$, we introduce the local connection forms $A_{\rm
 up/down}^+$ \cite{Nakahara} which are def\/ined through
\begin{gather*}
d |v_{\rm up}^+\rangle = |v_{\rm up}^+\rangle A_{\rm up}^+, \qquad d |v_{\rm down}^+\rangle = |v_{\rm down}^+\rangle A_{\rm down}^+,
\end{gather*}and related by
\begin{gather*}
A_{\rm up}^+ = A_{\rm down}^+ + \eta^{-1} d\eta.
\end{gather*}Since $\eta^{-1}d\eta$ is closed, the curvature form $F^+$ is globally def\/ined through
\begin{gather*}
F_{\rm up}^+ = d A_{\rm up}^+ = d A_{\rm down}^+ = F_{\rm down}^+.
\end{gather*}Finally, the Chern number is calculated by integrating the curvature $F^+$ by the use of the Stokes theorem,
\begin{gather*}
\int_{S^2} F^+ = \int_{S_-^2} dA_{\rm up}^+ + \int_{S_+^2} dA_{\rm down}^+ =\int_{-C} A_{\rm up}^+ + \int_{C} A_{\rm down}^+ \\
\hphantom{\int_{S^2} F^+}{} = \int_{C} \left(A_{\rm down}^+ - A_{\rm up}^+\right) = - \int_C \eta^{-1} d\eta = 2\pi i,
\end{gather*}
where $S_+^2$ and $S_-^2$ are the north and the south hemispheres, respectively, and $C$ is the equator with orientation in keeping with the natural orientation of $S_+^2$. Hence, the Chern number of the eigenline bundle associated with $\mu^+$ for $-1<A<1$ is
\begin{gather*}
\frac{i}{2\pi} \int_{S^2} F^+ = -1.
\end{gather*}The Chern number of the eigenline bundle associated with $\mu^-$ is $+1$ accordingly.

\subsubsection[Chern numbers for $H_{\text{\rm semi-quantum}}$ with $d=0$, $\delta=-1$]{Chern numbers for $\boldsymbol{H_{\text{\rm\bf semi-quantum}}}$ with $\boldsymbol{d=0}$, $\boldsymbol{\delta=-1}$}

In case of $\delta <0$, the Chern numbers of the eigenline bundle associated with $\mu^\pm$ for $-1 < A < 1$ are inversed in sign respectively. The calculation for $\delta=-1$ runs in parallel to the case of $\delta=1$, while the exceptional points $\pm\mathbf{e}_3$ given in table~(\ref{TableExP1}) are interchanged into $\mp\mathbf{e}_3$, respectively. We denote by $\mathrm{Ch}^\pm$ the Chern numbers associated with $\mu^\pm$. Then the above results are summed up to say that the Chern numbers in the interval, $-1<A<1$, of the control parameter are given by $\mathrm{Ch}^\pm=\mp1$ for $\delta>0$ and $\mathrm{Ch}^\pm=\pm1$ for $\delta<0$. These Chern numbers can be linked to the number of energy levels in the bands for the quantum Hamiltonian (\ref{L+L-}) with the intermediate parameter values $-1<A<1$, where $L$ is normalized to be $L=1$ and $d=0$ and $\delta=\pm1$. To make the correspondence easy to see, we assign the upper and lower bands by the sign $\pm$, and denote the number of energy levels in each band by $N^\pm$. Then one has $N^\pm = 2L+1-\mathrm{Ch}^\pm$ for $-1<A<1$, independently of the sign of $\delta$. This relation is valid also for $A$ with $|A|>1$, since $\mathrm{Ch}^\pm=0$ and $N^\pm=2L+1$ for $|A|>1$. Hence, the relation
\begin{gather} \label{N_Chern_relation}
 N^\pm=2L+1-\mathrm{Ch}^\pm
\end{gather}
holds for $A\neq \pm1$, independently of the sign of $\delta$. This relation means that the integers from the analysis of the quantum system are evaluated by using topological invariants from the analysis of the semi-quantum system. On account of topological nature, the present relation between the Chern numbers and the numbers of energy levels in the energy bands may be extended to be valid for arbitrary phenomenological and control parameters and even for an arbitrary number of bands, $S\ll L$, as long as the bands are isolated and there are no degeneracy points of eigenvalues so that the Chern numbers are def\/ined.

\subsection[Chern numbers for $H_{\text{\rm semi-quantum}}$ with $d=1$, $\delta=0$]{Chern numbers for $\boldsymbol{H_{\text{\rm\bf semi-quantum}}}$ with $\boldsymbol{d=1}$, $\boldsymbol{\delta=0}$}
The eigenvalues of $H_{\text{\rm semi-quantum}}$ with $d=1$, $\delta=0$ are given by
\begin{gather*}
\lambda^\pm = \pm \sqrt{ \big(A + x_3^2\big)^2 + |\gamma|^2 \big(x_1^2+x_2^2\big) }.
\end{gather*}This implies that a degeneracy in the eigenvalues occurs for $\gamma\neq0$ if and only if
\begin{gather*}
x_1=x_2=0, \qquad A+x_3^2 = 0.
\end{gather*}If $x_1=x_2=0$ then $x_3^2=1$ and the degeneracy occurs when $A=-1$, at $\pm \mathbf{e}_3$, the north and the south poles of the unit
sphere. The eigenline bundles are then def\/ined for $A\neq -1$.

In a way similar to the previous subsection, we f\/ind the normalized eigenvector associated with $\lambda^+$ in two ways,
\begin{alignat*}{3}
& |u_{\rm up}^+\rangle = \frac{1}{{N}_{\rm up}^+} \left( \begin{matrix}\bar{\gamma}(x_1-ix_2)\\
 \lambda_+ - \big(A+x_3^2\big) \end{matrix}\right), \qquad && {N}_{\rm up}^+ = \sqrt{ 2\lambda_+\big(\lambda_+ - \big(A+x_3^2\big)\big) }, & \\
& |u_{\rm down}^+\rangle = \frac{1}{{N}_{\rm down}^+} \left( \begin{matrix}\lambda_+ + A+x_3^2 \\
 \gamma(x_1+ix_2)\end{matrix}\right), \qquad && {N}_{\rm down}^+ = \sqrt{ 2\lambda_+\big(\lambda_+ + A+x_3^2\big) }.&
\end{alignat*}
The exceptional points for $|u_{\rm up}^+\rangle$ and $|u_{\rm down}^+\rangle$ are determined by the equations $\{x_1=x_2=0$, $\lambda_+ - (A+x_3^2) = 0\}$ and $\{x_1=x_2=0,\, \lambda_+ + A+x_3^2 = 0\}$, respectively, and then listed in table: %~(\ref{Table2}):
\begin{gather} \label{Table2}
\begin{array}{c|c|c|c}
 & A<-1 & A=-1 & A>-1 \\ \hline
{\rm except.\ pts.\ for\ } |u_{\rm up}^+\rangle & {\rm no} &
\left(\begin{matrix} \pm \mathbf{e}_3 \\ {\rm deg.\ pts.}
\end{matrix} \right) & \pm \mathbf{e}_3 \\ \hline
{\rm except.\ pts.\ for\ } |u_{\rm down}^+\rangle & \pm \mathbf{e}_3 &
\left(\begin{matrix} \pm \mathbf{e}_3 \\ {\rm deg.\ pts.}
\end{matrix} \right) & {\rm no}
\end{array}
\end{gather}
Table~(\ref{Table2}) implies that the eigenvectors $|u_{\rm up}^+\rangle$ and $|u_{\rm down}^+\rangle$ are globally def\/ined for $A<-1$ and for $A>-1$, respectively. Hence the eigenline bundle associated with $\lambda^+$ is trivial for $A\neq -1$. In a similar manner, the eigenline bundle associated with $\lambda^-$ proves to be trivial for $A\neq -1$. It then turns out that the Chern numbers of the eigenline bundles associated with $\lambda^\pm$ are zero for $A\neq-1$.

\begin{figure}[t]\centering
 \includegraphics[scale=0.41]{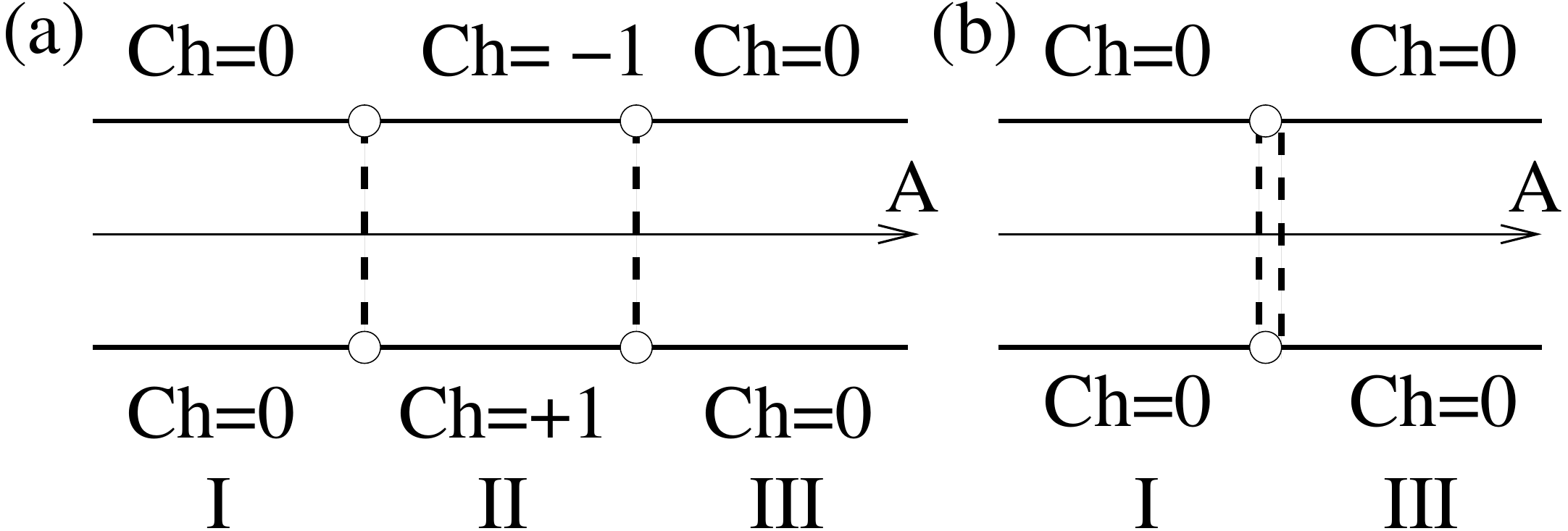}
\caption{Schematic representation of the evolution of the two eigenline bundles of $H_{\text{\rm semi-quantum}}$ Hamiltonian ($S=\frac{1}{2}$). (a) Case of $\delta > 0$. (b) Case of $\delta = 0$. Note that the two local delta-Cherns associated with degeneracies at south and north poles remain non-zero, but they are summed up to give a global delta-Chern with zero value. For $\delta < 0$ the Chern numbers $\mathrm{Ch}=\pm1$ should be interchanged. For further comments see text.} \label{SchemChern2}
\end{figure}

Fig.~\ref{SchemChern2} shows schematically the evolution of two bands along with the variation of the control parameter~$A$, which are represented by solid lines together with the associated Chern numbers. The dashed lines symbolize the formation of the degeneracy points between the two eigenvalues of the semi-quantum Hamiltonian.

\looseness=1 Fig.~\ref{SchemChern2}(a) corresponds to the Hamiltonian with $\delta> 0$. The dif\/ference of $A$ values for the two degeneracy points is proportional to the $\delta$ parameter. From table~(\ref{TableExP1}) and Fig.~\ref{SchemChern2}(a), we see that when the value of $A$ passes the critical value $A=-1$, the Chern number of the eigenline bundle associated with $\mu^+$ changes from $0$ to $-1$ because of the existence of the degeneracy point $+\boldsymbol{e}_3$, and when $A$ passes $A=+1$, the Chern number changes from $-1$ to $0$ because of the existence of the degeneracy point $-\boldsymbol{e}_3$. We here def\/ine the local delta-Chern assigned to a degeneracy point, when the parameter takes a critical value, to be the contribution to the change in the Chern number in the positive direction of the parameter $A$ and the delta-Chern to be the sum of local delta-Cherns from all degeneracy points. Then, for $\mu^+$ we may assign the local delta-Chern $-1$ to $+\boldsymbol{e}_3$ when $A=-1$ and the local delta-Chern $+1$ \smash{($=0-(-1)$)} to $-\boldsymbol{e}_3$ when $A=+1$. Since the degeneracy point is $+\boldsymbol{e}_3$ only when $A=-1$ and since the degeneracy point is $-\boldsymbol{e}_3$ only when $A=+1$, the delta-Cherns when $A=\mp 1$ are $\mp 1$, respectively. The total variation of the Chern number for the whole range of $A$ is the sum \smash{$(-1)+1=0$}.

Fig.~\ref{SchemChern2}(b) corresponds to the semi-quantum Hamiltonian with $\delta=0$. In this case, there appear two isolated degeneracy points $\pm\boldsymbol{e}_3$ simultaneously when $A=-1$. The local contribution to the modif\/ication of Chern numbers from the degeneracy points $\pm\boldsymbol{e}_3$ (see (\ref{Table2})) have the same absolute values with opposite signs, so that the sum of the delta-Cherns is $1+(-1)=0$, and hence no change is found in the Chern number, as is seen in Fig.~\ref{SchemChern2}(b).

Figs.~\ref{Z2axialDelta} and \ref{SchemChern2} are in mutual correspondence. In Fig.~\ref{Z2axialDelta}(b) and in Fig.~\ref{SchemChern2}(b), we observe that the band rearrangement and the modif\/ication of Chern number take place at a critical value of $A$ (though the number of levels belonging to each band is invariant and though the total change of the Chern number is $+1+(-1)=0$). In Fig.~\ref{Z2axialDelta}(a) and Fig.~\ref{SchemChern2}(a), we see that there is an intermediate region~II of $A$ values for which in the full quantum model [Fig.~\ref{Z2axialDelta}(a)] the quantum energy bands consist of a dif\/ferent number of energy levels and for semi-quantum model [Fig.~\ref{SchemChern2}(a)] the Chern numbers for the eigenline bundles associated with $\mu^+$ and $\mu^-$ are non-zero.

\subsection[Generalization to $N$-bands]{Generalization to $\boldsymbol{N}$-bands}\label{S33}

In correspondence to the band inversion treated quantum mechanically in Section~\ref{S22}, the semi-quantum version of the Hamiltonian~(\ref{Hamiltonian_ref}) in the case of an arbitrary number of bands, i.e., in the case of an arbitrary pseudo-spin number $S$, is expected to exhibit a corresponding inversion of the band structure when the system of eigenvalues is compared in the two limits as $A\rightarrow \infty$ and as $A\rightarrow -\infty$. This is anticipated from the expression in the semi-quantum Hamiltonian
\begin{gather} \label{sqh}
 H_{\text{\rm semi-quantum}}=2S_z\big(A+\delta x_3+d x_3^2\big) +\gamma S_- (x_1+ix_2) + \bar{\gamma} S_+ (x_1-ix_2).
\end{gather}

\subsubsection{Numbers of energy levels and Chern numbers}

The relation~(\ref{N_Chern_relation}) between the Chern number and the number of energy levels for a two-band system can be generalized to the $N$-band case with $N=2S+1$. For an $N$-band system, a semi-quantum Hamiltonian is generically described as a Hermitian $N\times N$ matrix. The degeneracies of the eigenvalues of such a matrix are of codimension~3. Each matrix element of the semi-quantum Hamiltonian is def\/ined on the two-dimensional base space and depends furthermore on the external control parameter~$A$. As a consequence, degeneracy is generically allowed between two neighboring eigenvalues only. Near each such degeneracy, the relation between the number of quantum states in the band and the Chern number for the eigenline bundle of the semi-quantum Hamiltonian is valid. Consequently, the relation~(\ref{N_Chern_relation}) remains valid for an arbitrary sequence of rearrangements associated with the formation of degeneracy points between two bands.

However, the Hamiltonian~(\ref{sqh}) does not exhibit a typical rearrangement of the band structure under the variation of one control parameter, which is already stated above. The present Hamiltonian admits a multiple degeneracy among several eigenvalues simultaneously, but by a~small deformation of the Hamiltonian (for example, by adding terms of higher degrees in $S_{\alpha}$), it is possible to unfold the multiple degeneracy into a sequence of generic degeneracies between only two neighboring energy levels. After such a deformation, we can apply again a sequence of relations~(\ref{N_Chern_relation}). With this in mind, we can keep only linear in $S_{\alpha}$ terms introduced in~(\ref{sqh}) and study the evolution of the Chern numbers of the eigenline bundles of the semi-quantum Hamiltonian against the control parameter~$A$.

\subsubsection[Chern numbers for an arbitrary number $N=2S+1$ of bands]{Chern numbers for an arbitrary number $\boldsymbol{N=2S+1}$ of bands}
The corresponding semi-quantum Hamiltonian~(\ref{sqh}) has the following tridiagonal form for arbitrary $S$:
\begin{gather} \label{add4}
 H_{\text{\rm s-q}}(\mathbf{x};A) =
 \left( \begin{matrix}
 H_{1,1}(\mathbf{x};A) & H_{2,1}(\mathbf{x};A)^* & \cdots & 0 & 0 \\
 H_{2,1}(\mathbf{x};A) & H_{2,2}(\mathbf{x};A) & \cdots & 0 & 0 \\
 \vdots & \vdots & \ddots & \vdots & \vdots \\
 0 & 0 & \cdots & H_{2S,2S}(\mathbf{x};A) & H_{2S+1,2S}(\mathbf{x};A)^* \\
 0 & 0 & \cdots & H_{2S+1,2S}(\mathbf{x};A) & H_{2S+1,2S+1}(\mathbf{x};A) \\
 \end{matrix} \right),\!\!\!\!
\end{gather}
where
\begin{gather*}
\begin{split} & H_{i,i}(\mathbf{x};A) = 2 (S+1-i )f(x_3;A), \\
 & H_{i+1,i}(\mathbf{x};A) = \sqrt{S (S+1 )- (S+1-i ) (S-i )}\gamma (x_1+ix_2).
\end{split}
\end{gather*}
The Hamiltonian~(\ref{add4}) has degeneracy points of eigenvalues only at $\mathbf{x}=\pm\mathbf{e}_3$, the north and the south poles of the unit sphere, owing to the rotational symmetry. Evaluated at~$\pm\mathbf{e}_3$, the Hamiltonian is expressed as
\begin{gather*}
H_{\rm s-q}(\pm \mathbf{e}_3; A) = 2f(x_3;A)\, {\rm diag}(S, (S-1), \dots, -(S-1), S), \qquad x_3=\pm 1.
\end{gather*}The present form of Hamiltonian immediately gives the eigenvalues evaluated at $\pm \mathbf{e}_3$, from which it turns out that the degeneracy of eigenvalues is $N$-fold with $N=2S+1$ and the $N$-fold degeneracy in eigenvalues occurs at $\mathbf{e}_3$ and at $-\mathbf{e}_3$ when $A=-(\delta+d)$ and when $A=\delta-d$, respectively.

The whole range of the control parameter $A$ is broken up into three disjoint open subsets, the iso-Chern domains, on which the Chern numbers of the eigenline bundles associated with the eigenvalues of Hamiltonian~(\ref{sqh}) are constant. The eigenline bundles are trivial (so that the Chern numbers for respective bands are zero) for the two extremal intervals of the control parameter, which are denoted by the domain~I with $A<-d-|\delta|$ and the domain~III with $A>-d+|\delta|$. The eigenline bundles are non-trivial in the intermediate domain II with $-d-|\delta|<A<-d+|\delta|$. The width of the intermediate domain~II is $2|\delta|$, independent of the parame\-ter~$d$. The parameter $d$ shifts only the position of the domain~II in the whole range of the control parameter~$A$. Hence, in evaluating the Chern number of an eigenline bundle in the domain~II, one may choose $d=0$ without changing the value of the Chern number. Then, the domain II is shifted to $-|\delta|<A<|\delta|$. In what follows, we f\/irst consider the case $\delta>0$. Since the value of the Chern number is independent of $\delta$, we may choose $\delta=1$ and take the point $A=0$ to evaluate the Chern number. Furthermore, the Chern number is independent of~$\gamma$, so that one can set $\gamma=1$. Then, the semi-quantum Hamiltonian~(\ref{sqh}) reduces to
\begin{gather*}
 H = 2 ( x_1 S_x + x_2 S_y + x_3 S_z ).
\end{gather*}From the representation theory of ${\rm SU}(2)$, the pseudo-spin operator~$S_z$ receives the transformation under the action by $D(g)$,
\begin{gather*}
 D(g) 2S_z D(g)^\dagger = 2\big[\cos\theta S_z+\sin\theta(S_x\cos \phi+S_y \sin \phi) \big],
\end{gather*}where $D(g)$ denotes the unitary representation matrix
\begin{gather*}
 D(g)=e^{-i\phi S_z}e^{-i\theta S_y}e^{-i\psi S_z},
\end{gather*}and where $\left(\theta,\phi,\psi\right)$ are the Euler angles. This equation implies that the unitary matrix $D(g)$ diagonalizes the Hamiltonian $H$ def\/ined on the sphere $S^2$,
\begin{gather*}
 H = 2D(g)^\dagger S_z D(g).
\end{gather*}This also means that the eigenvalues of $H$ are given by $2r$ with $|r|\leq S$.

The eigenvector associated with the eigenvalue $2r$ of $H$ is given by
\begin{gather*}
 D(g)|S,r\rangle =e^{-i\phi S_z} e^{-i\theta S_y} e^{-i\psi S_z}|S,r\rangle,
\end{gather*}for any value of $\psi$. In order to describe the eigenvectors as vector-functions on the two-sphere, we need to choose $\psi$ as a function on the two-sphere. An easy way to do so is to take $\psi=\pm \phi$. If $\psi=-\phi$, we have
\begin{gather*}
 e^{-i\phi S_z} e^{-i\theta S_y} e^{i\phi S_z}|S,r\rangle. \end{gather*}If we let $\theta$ tend to 0, this eigenvector tends to $|S,r\rangle$, which means that the limit eigenvector is uniquely determined irrespective of $\phi$, so that this eigenvector is def\/ined at the north pole of the sphere. In contrast to this, if we let $\theta \to \pi$, the limit is not determined irrespective of $\phi$, $e^{-i\phi S_z} e^{-i\pi S_y} e^{i\phi S_z}|S,r\rangle$, so that the eigenvector of the present form is not def\/ined at the south pole. Put another way, the south pole is an exceptional point. Thus, we have obtained an expression of the eigenvector associated with the eigenvalue $2r$,
\begin{gather*}
|u_r(\boldsymbol{x})_+\rangle = e^{-i\phi S_z} e^{-i\theta S_y} e^{i\phi S_z}|S,r\rangle, \qquad U_+=S^2-\{-\boldsymbol{e}_3\}.
\end{gather*}

Another expression of the eigenvalue is given by setting $\psi=\phi$,
\begin{gather*}
 e^{-i\phi S_z} e^{-i\theta S_y} e^{-i\phi S_z}|S,r\rangle.
\end{gather*}If we let $\theta$ tend to 0, we have $e^{-2i\phi S_z}|S,r\rangle$, which means that the limit is not unique, so that this eigenvalue is not def\/ined at the north pole. Contrary to this, if we let $\theta\to \pi$, we have a unique limit $e^{-i\phi S_z} e^{-i\pi S_y} e^{-i\phi S_z}|S,r\rangle = e^{-i\pi S_y}|S,r\rangle$. Put another way, this eigenvector is def\/ined at the south pole. Thus, we have found another expression of the eigenvector,
\begin{gather*}
|u_r(\boldsymbol{x})_-\rangle= e^{-i\phi S_z} e^{-i\theta S_y} e^{-i\phi S_z}|S,r\rangle, \qquad U_-=S^2-\{\boldsymbol{e}_3\}.
\end{gather*}
By using the expression of the matrix $e^{-i\phi S_z} e^{-i\theta S_y} e^{-i\phi S_z}$ with respect to the canonical basis $|S,r\rangle$, we f\/ind that the eigenvectors expressed in two ways are related by
\begin{gather*}
e^{-2ir\phi}|u_r(\boldsymbol{x})_+\rangle = |u_r(\boldsymbol{x})_-\rangle \qquad {\rm on} \quad U_+\cap U_-.
\end{gather*}

The local connection forms $A_{\pm}^{(r)}$ on $U_{\pm}$ are def\/ined to be
\begin{gather*}
A_{\pm}^{(r)}= \langle u_r(\boldsymbol{x})_{\pm}|d|u_r(\boldsymbol{x})_{\pm}\rangle \qquad {\rm on} \quad U_{\pm},
\end{gather*}where the superscript $(r)$ indicates that these forms are assigned to the eigenline bundle associated with the eigenvalue $2r$ of $H$. These local connection forms are shown to be related
\begin{gather*}
 A^{(r)}_+ -2ir\, d\phi = A^{(r)}_- \qquad {\rm on} \quad U_+\cap U_-.
\end{gather*}The local curvature forms are def\/ined to be{\samepage
\begin{gather*}
 F^{(r)}_{\pm}=dA^{(r)}_{\pm} \qquad {\rm on} \quad U_{\pm},
\end{gather*}which are put together to def\/ine the global curvature form $F^{(r)}$.}

Now the calculation of the Chern number is straightforward, which runs as follows:
\begin{gather*}
 \int_{S^2} F^{(r)} = \int_{S^2_+} dA^{(r)}_+ + \int_{S^2_-} dA^{(r)}_- =\int_C A^{(r)}_+ + \int_{-C}A^{(r)}_- \\
 \hphantom{\int_{S^2} F^{(r)}}{} = \int_C \big(A^{(r)}_+-A^{(r)}_-\big)= 2ir \int_C d\phi = 4\pi ir.
\end{gather*}Thus, the Chern number of the eigenline bundle associated with the eigenvalue $2r$ is
\begin{gather*}
 \frac{i}{2\pi}\int_{S^2}F^{(r)} =-2r.
\end{gather*}

To sum up the discussion on the Chern numbers for an arbitrary number of bands, the $N$-dimensional vector of Chern numbers is subject to the following changes for $\delta>0$:
\begin{gather} \label{chern_changes}
\underbrace{\left( \begin{matrix} 0\\0\\\vdots\\0\\0 \end{matrix} \right)}_{\mathrm{domain\ I}} \xrightarrow{\begin{matrix} \text{\footnotesize wall\ between}\\ \text{\footnotesize domains\ I\ and\ II}\end{matrix}}
\underbrace{\left( \begin{matrix} -2S\\-2(S-1)\\\vdots\\2(S-1)\\2S \end{matrix} \right)}_{\mathrm{domain\ II}} \xrightarrow{\begin{matrix} \text{\footnotesize wall\ between}\\ \text{\footnotesize domains\ II\ and\ III}\end{matrix}}
\underbrace{\left( \begin{matrix} 0\\0\\\vdots\\0\\0 \end{matrix} \right)}_{\mathrm{domain\ III}},
\end{gather}
where the Chern numbers are ordered in such a manner that the top ones ($0$, $-2S$, and $0$ in their respective columns) are associated with the highest energy eigenvalue and the second top ones with the second highest eigenvalue, downward to the bottom ones ($0$, $2S$, $0$ in the three columns) with the lowest eigenvalue.

If $\delta$ is negative, then the same calculation leads to the conclusion that the Chern numbers in the second column change sign.

\section{Completely classical version and energy-momentum map}\label{S_4}
The redistribution of energy levels bears the marks even in the completely classical limit of the Hamiltonian~(\ref{Hamiltonian_ref}). In this limit, all dynamical variables are treated as classical variables. For notational simplicity, we use the same symbols $S_{\alpha}$ and $L_{\alpha}$ as classical variables. However, they are assumed to be subject to the conditions that $ |\mathbf{L} |^2=L_x^2+L_y^2+L_z^2 = {\rm const}$ and
$|\mathbf{S}|^2=S_x^2+S_y^2+S_z^2={\rm const}$. Then, the phase space for the classical system is the direct product, $S^2\times
S^2$, of two two-dimensional spheres with the radii given by the constants~$ |\mathbf{L} |$ and~$ |\mathbf{S} |$, respectively. The group ${\rm SO}(2)$ acts on $S^2\times S^2$ in the manner such that the ${\rm SO}(2)$ group action is a simultaneous rotation about the $z$-axis of each factor space~$S^2$.\footnote{For more general weighted action and its relevance to physical examples, see \cite{ RevModPhys, Hansen, IwaiZhTheorChemAcc, Nekhoroshev}.} The phase space $S^2\times S^2$ is endowed with the canonical symplectic structure which is alternatively described in terms of the Poisson brackets among $S_{a}$ and $L_{b}$, $a,b\in \{x,y,z\}$,
\begin{gather*}
 \{S_{a},S_{b}\}=\sum \varepsilon_{abc} S_{c}, \qquad \{L_{a},L_{b}\} =\sum \varepsilon_{abc} L_{c}, \qquad \{S_a,L_b\}=0.
\end{gather*}
The corresponding classical Hamiltonian is called $H_\mathrm{classical}$. By denoting the real and imaginary parts of $\gamma$ by $\gamma_r$ and $\gamma_i$, respectively and by introducing the ${\rm SO}(2)$-invariant polynomials
\begin{gather*}
 \tau=S_xL_x + S_yL_y, \qquad \sigma=S_xL_y-S_yL_x,
\end{gather*}
the present Hamiltonian is rewritten as
\begin{gather} \label{Hclass}
H_\mathrm{classical} = 2S_z\big(A+\delta L_z +d L_z^2\big) + 2 \gamma_r \tau - 2 \gamma_i \sigma.
\end{gather}
On account of the ${\rm SO}(2)$ symmetry of $H_\mathrm{classical}$, the quantity $J_z=L_z+S_z$ is an integral of motion, $\{J_z, H_\mathrm{classical}\}=0$, so that the present Hamiltonian system is completely integrable. We note that dynamical models with the $S^2\times S^2$ phase space and with a ${\rm SO}(2)$ invariance symmetry appear in molecular problems in quite dif\/ferent contexts, such as the coupling of angular momenta \cite{Grondin, PhysLettMono}, the hydrogen atom in weak external f\/ields \cite{RevModPhys, Ferer}, the rotational structure of bending modes in quasi-linear molecules~\cite{Child1, Winnewisser}, or the internal structure of vibrational polyads formed by two doubly degenerate vibrational modes in linear molecules, like acetylene $\mathrm{C_2H_2}$~\cite{Herman, Kellman}.

\begin{figure}[t]\centering \includegraphics[scale=0.76]{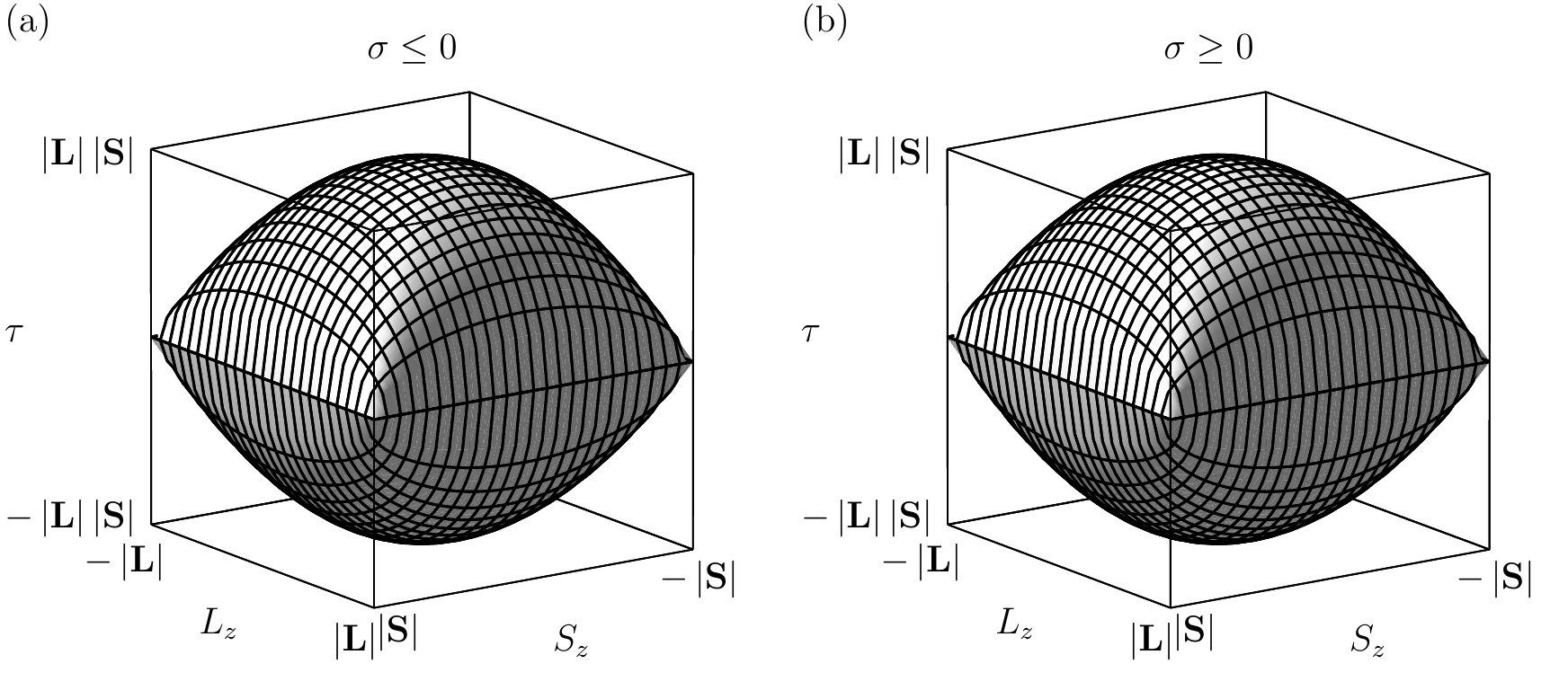}
\caption{Space of orbits of the ${\rm SO}(2)$ group action on $S^2\times S^2$. The boundaries of the respective solid bodies are glued together at corresponding points.}\label{calculs_Z2model_020}
\end{figure}

The reduction of such dynamical systems by the ${\rm SO}(2)$ symmetry has been discussed on several occasions~\cite{Cushman, RevModPhys, PhysLettMono}. Our discussion below follows essentially~\cite{PhysLettMono}. To describe the space of orbits of the ${\rm SO}(2)$ group action on the phase space $S^2\times S^2$, we use the ${\rm SO}(2)$-invariant polynomials, $L_z$, $S_z$, $\tau$, and $\sigma$, among which the polynomials $L_z$, $S_z$, $\tau$ are algebraically independent and any linear combination of these polynomials can be equally used, and the polynomial $\sigma$ is an auxiliary polynomial, i.e., it is only linearly independent and related by the following syzygy to the basic independent polynomials~\cite{MichelPhysRep}:
\begin{gather} \label{Syzygy}
 \sigma^2=\big( |\mathbf{S} |^2-S_z^2\big) \big( |\mathbf{L} |^2-L_z^2\big)-\tau^2.
\end{gather}
Since $\sigma^2\geq 0$, the right-hand side of equation~(\ref{Syzygy}) determines in the space $\{ S_z, L_z, \tau\}$ a f\/inite volume representing the space of orbits through
\begin{gather*}
\tau^2 \leq \big(|\mathbf{S}|^2-S_z^2\big) \big(|\mathbf{L}|^2-L_z^2\big).
\end{gather*}
Every point on the boundary of this space of orbits corresponds to one orbit with $\sigma=0$, whereas every internal point corresponds to two orbits distinguished by the sign of~$\sigma$. That is why the graphical representation of the space of orbits in Fig.~\ref{calculs_Z2model_020} consists of two bodies and the respective points in the boundary of these two bodies should be identif\/ied. There are four singular points at $\{ S_z=\pm |\mathbf{S}|,\, L_z=\pm |\mathbf{L}| \}$. These points are isolated orbits of the ${\rm SO}(2)$ action with the stabilizer being the ${\rm SO}(2)$ group itself.

\begin{figure}[t]\centering
 \includegraphics[scale=0.76]{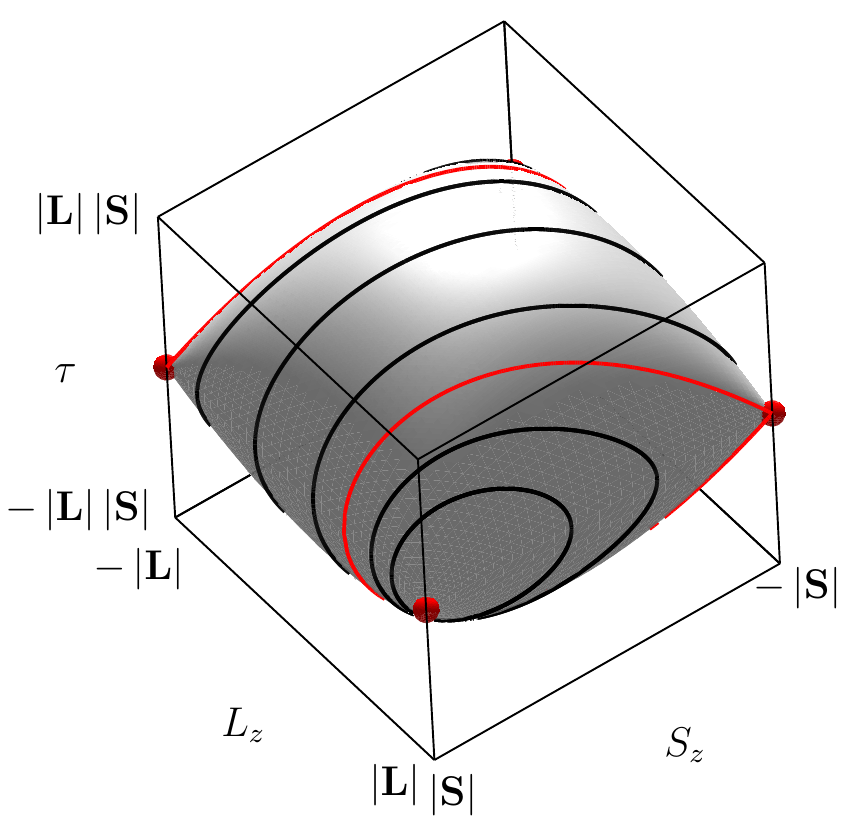}
\caption{Space of orbits sliced by $J_z=\mathrm{const}$ planes. Black curves: boundary curves of the intersection of the space of orbit (with $\sigma\geq 0$ of $\sigma\neq 0$) with $J_z={\rm const}$ planes, which contains only regular points. Red curves: boundary curves of the intersection of the space of orbits with $J_z={\rm const}$ planes, which contains a singular point.}\label{pillow_sliced}
\end{figure}

\begin{figure}[t]\centering
 \includegraphics[scale=0.76]{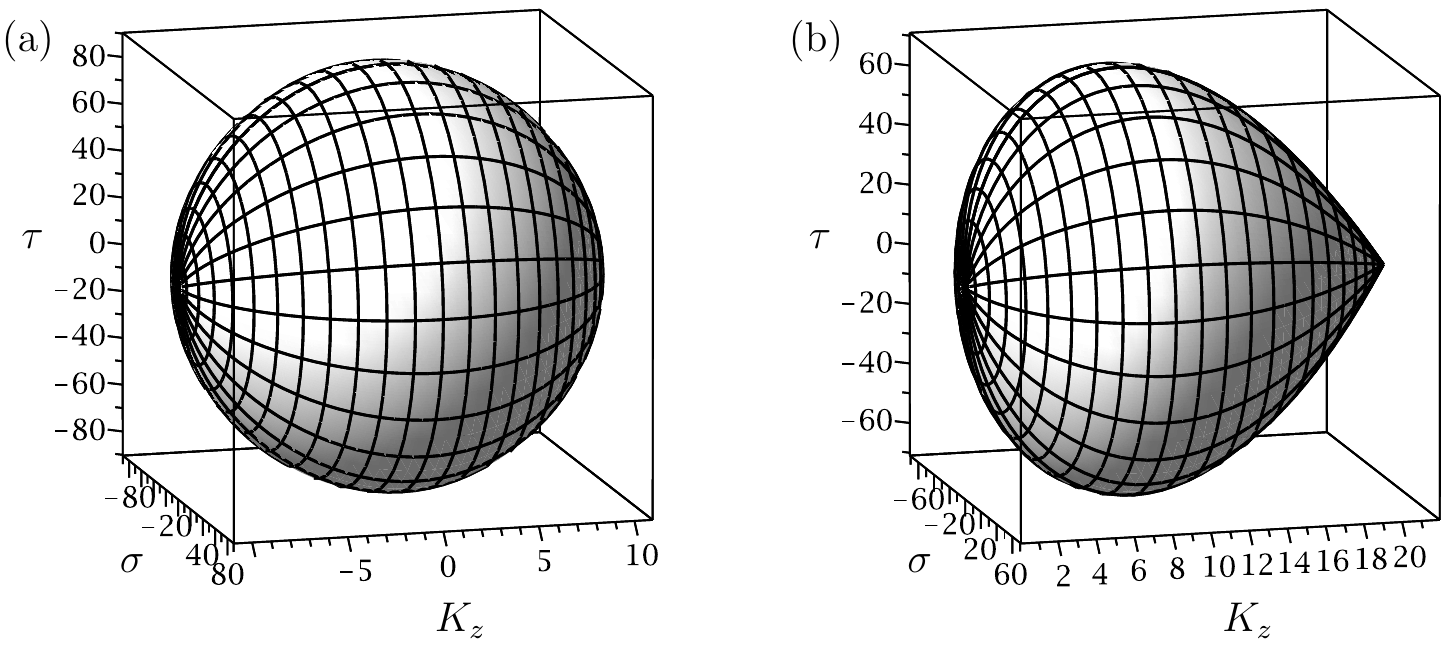}
\caption{(a) Regular reduced phase space for $J_z=0$. (b) Singular reduced phase space for $J_z=|\mathbf{S}|-|\mathbf{L}|$.} \label{reduced_phase_space}
\end{figure}

\looseness=-1 The reduced phase space assigned by the condition $J_z=\mathrm{const}$ can be geometrically obtained by cutting each part of the space of orbits by the $J_z=\mathrm{const}$ plane as in Fig.~\ref{pillow_sliced} and by identifying respective points on the boundary of the two obtained sections. The result of such a~construction is generically a topological sphere (see Fig.~\ref{reduced_phase_space}(a)). In particular, for $J_z=\pm(|\mathbf{L}|+|\mathbf{S}|)$, the reduced phase space becomes a singular point and for $J_z=\pm(|\mathbf{L}|-|\mathbf{S}|)$ it becomes a sphere with a singular point on it (see Fig.~\ref{reduced_phase_space}(b)). Using the ${\rm SO}(2)$-invariant polynomials $\tau$, $K_z=S_z-L_z$ and $\sigma$ as coordinates instead of $L_z$, $S_z$, and $\tau$, we can describe the reduced phase space as a topological sphere.

The reduced phase space is the base space of a f\/iber bundle whose f\/iber is a ${\rm SO}(2)$ orbit which is generically a circle. The intersection of this f\/iber bundle with an energy $H=\mathrm{const}$ surface gives generically a two-dimensional torus. When this section passes through a critical value, i.e., through one of the points with $\{ L_z=|\mathbf{L}|,\, S_z=-|\mathbf{S}| \}$ or $\{L_z=-|\mathbf{L}|,\, S_z=|\mathbf{S}| \}$, the intersection in question becomes a singly pinched torus instead of a~regular two-dimensional torus, since one circle is shrunk to a~point. This singly pinched torus corresponds to a critical value of the energy-momentum map, often called a monodromy point.

In the case of $|\mathbf{L}|=|\mathbf{S}|$, the $J_z=0$ section of the space of orbits is a sphere with two singular points and a doubly pinched tori can appear for some Hamiltonians. This is the case for ef\/fective Hamiltonians describing the internal structure of the perturbed hydrogen atom shell (see~\cite{RevModPhys} and references therein).

In order to see the correspondence between the redistribution of energy levels occurring in the quantum system under the variation of the control parameter $A$ and the qualitative modif\/ications to appear in the corresponding completely classical system, we study the evolution of the image of the energy-momentum map for the completely classical integrable system with Hamiltonian~(\ref{Hclass}) under the variation of $A$.

\begin{figure}[t]\centering
 \includegraphics[scale=0.80]{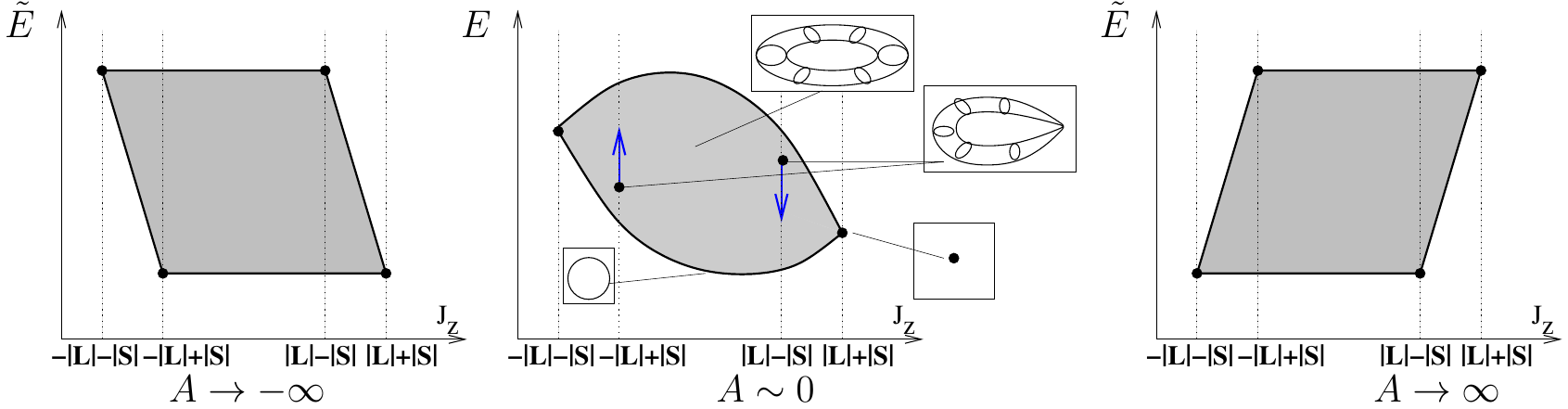}
\caption{Image of the energy-momentum map for Hamiltonian $H_\mathrm{classical}$ (\ref{Hclass}) with $\delta\approx 0$, $d\approx 0$, $\gamma_i \approx 0$. The blue arrows show the displacement of the critical values with increasing $A$. (a) Limit $A \rightarrow -\infty$. (b)~$A\sim 0$. (c) Limit $A \rightarrow \infty$.}\label{EnMomLim}
\end{figure}

The images of the energy-momentum map for the present Hamiltonian system with very big negative $A$ and very big positive $A$ are shown in the leftmost and the rightmost panels of Fig.~\ref{EnMomLim}, respectively. The middle panel shows the image of the energy-momentum map for an intermediate value of $A$. The shape of the image of the energy-momentum map for big $|A|$ follows immediately from the fact that for big $|A|$ the Hamiltonian can be approximated by $H_\mathrm{classical}\approx 2 A S_z$. There are four isolated critical values of the energy-momentum map, which are shown by black dots. For big $|A|$, those critical points belong to the boundary of the image. The inverse image of each of the regular points on the boundary under the energy-momentum map is a one-dimensional torus (circle) in the total phase space $S^2\times S^2$. For each of regular internal points of the image of the energy-momentum map, the inverse image is a regular two-dimensional torus in $S^2\times S^2$.

We now look into the evolution of the image of the energy-momentum map accompanying the variation in the control parameter $A$ from big negative to big positive values. The f\/igure in the leftmost panel of Fig.~\ref{EnMomLim} transforms to that in the rightmost panel of Fig.~\ref{EnMomLim}, taking a shape like that shown in the middle panel of Fig.~\ref{EnMomLim}. The axial symmetry of the problem is conserved during the evolution, and the critical values of the energy-momentum map keep belonging to respective sections with $J_z \in \{-|\mathbf{L}|-|\mathbf{S}|,\, -|\mathbf{L}|+|\mathbf{S}|,\, |\mathbf{L}|-|\mathbf{S}|,\, |\mathbf{L}|+|\mathbf{S}|\}$. For intermediate $A$ values, the image of the energy-momentum map may be more complicated than that shown in the middle panel of Fig.~\ref{EnMomLim} in the case of the appearance of additional critical values. However, as long as we are interested in the transition between $A\rightarrow -\infty$ and $A\rightarrow \infty$ limits, it is suf\/f\/icient to study the correlation between the two limits, so that we can adopt the simplest possible scenario for such a transition. For this reason, we can put $\delta=d=\gamma_i=0$ and keep $\gamma_r\neq0$ in order to avoid the collapse of the image of energy-momentum map for $A=0$, and the middle panel of Fig.~\ref{EnMomLim} is drawn with this selection of parameters. For such a simplif\/ied family of ef\/fective Hamiltonians depending on one control parameter $A$, isolated critical values appear as singular points within the domain of regular values. The inverse image of such an isolated critical value is a singly pinched torus in $S^2\times S^2$ \cite{Cushman}. The middle panel of Fig.~\ref{EnMomLim} schematically shows that two critical values move from one boundary of the image of the energy-momentum map to another boundary in opposite directions, accompanying the variation of $A$. We note in addition that the existence of singular values means also that the action-angle variables are not globally def\/ined and that the system exhibits Hamiltonian monodromy \cite{Duist}.

\looseness=1 A reason for the Hamiltonian $H_\mathrm{classical}$ with $\delta=0$ to be of special interest is that the corresponding quantum Hamiltonian with $\delta=0$ satisf\/ies the energy-ref\/lection symmetry condition. For the complete classical problem (\ref{Hclass}) with $\delta=0$, the image of the energy-momentum map has additional symmetry. If the point $(E,J_z)$ belongs to the image of the energy-momentum map, then the point $(-E,-J_z)$ also belongs to the image of the map, and the inverse images of these two points are equivalent. This is because in correspondence to the energy-ref\/lection symmetry transformation the classical variables $S_z, S_+$ and $S_-$ are subject to the transformation $S_z\mapsto -S_z$, $S_+\mapsto S_-$, and $S_-\mapsto S_+$ and the classical variables $L_z, L_+$, and $L_-$ to the transformation like (\ref{L variable transform}), so that the transformation $(E,J_z)\mapsto (-E,-J_z)$ follows imme\-dia\-tely.

Returning to the initial problem with non-zero phenomenological parameter values, we remark that the values of $A$ at which the critical value of the energy-momentum map passes from the boundary of the image to the inside of the image depend on the phenomenological parameters $d$, $\delta$, $\gamma$ and correspond to the Hamiltonian Hopf bifurcation~\cite{KEbook}. The topological character of the qualitative modif\/ications of the image of the energy-momentum map follows immediately from the dif\/ferent topology of the inverse images of the critical values at dif\/ferent values of the control parameter~$A$.

\begin{figure}[t]\centering
 \includegraphics[scale=0.45]{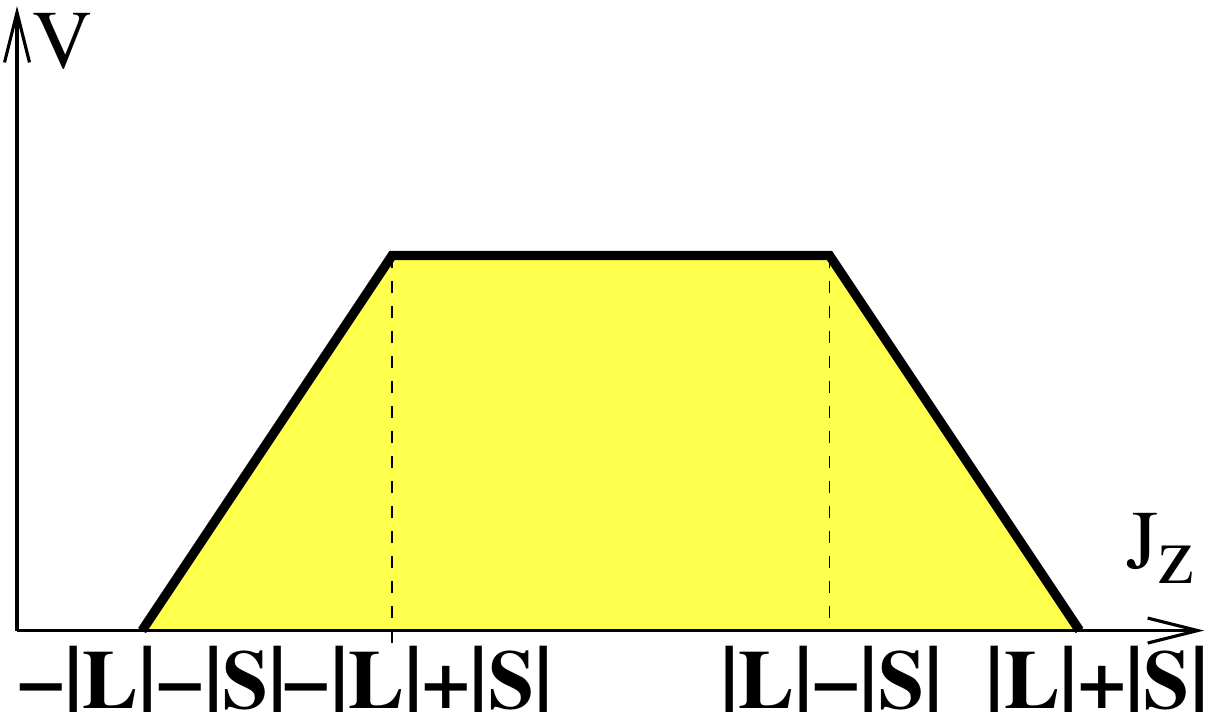}
\caption{Volume of the reduced phase space $V$ as function of the integral of motion $J_z$ for a classical dynamical system def\/ined over the $S^2\times S^2$ classical phase space in the presence of axial symmetry.}\label{DuistHFIG}
\end{figure}

A further observation on the present classical integrable dynamical system is obtained by applying the Duistermaat--Heckman theorem~\cite{DuistrH, Guillemin}, which says (in the concrete case we are studying) that the volume of the reduced phase space is a~piecewise linear function of the integral of motion. The discontinuity of the f\/irst derivative of the reduced volume with respect to the integral of the motion value occurs at $J_z$ values corresponding to critical values of the energy-momentum map, as is seen in Fig.~\ref{DuistHFIG}.

\section[Returning to the quantum picture: comparison with the semi-quantum and the classical point of views]{Returning to the quantum picture: comparison\\ with the semi-quantum and the classical point of views} \label{S_5}

So far we have independently studied quantum, semi-quantum, and classical Hamiltonians with an interest in the band rearrangements and the corresponding topological phase transitions. We now return to the quantum problem by passing from the image of the classical energy-momentum map of Section~\ref{S_4} to a representation of a joint spectrum of commuting quantum observables or the lattice of quantum states. This rather simple procedure applied to the complete classical model leads to results which helps the interpretation of the rearrangement of energy levels in terms of evolution of the energy-momentum map. To this end, it is instructive to look for the evolution of the lattice of the quantum states for low $S$ values, assuming the existence of $2S+1$ energy bands separated in the ideal case by energy gaps.

\begin{figure}[th!]\centering
 \includegraphics[scale=0.77]{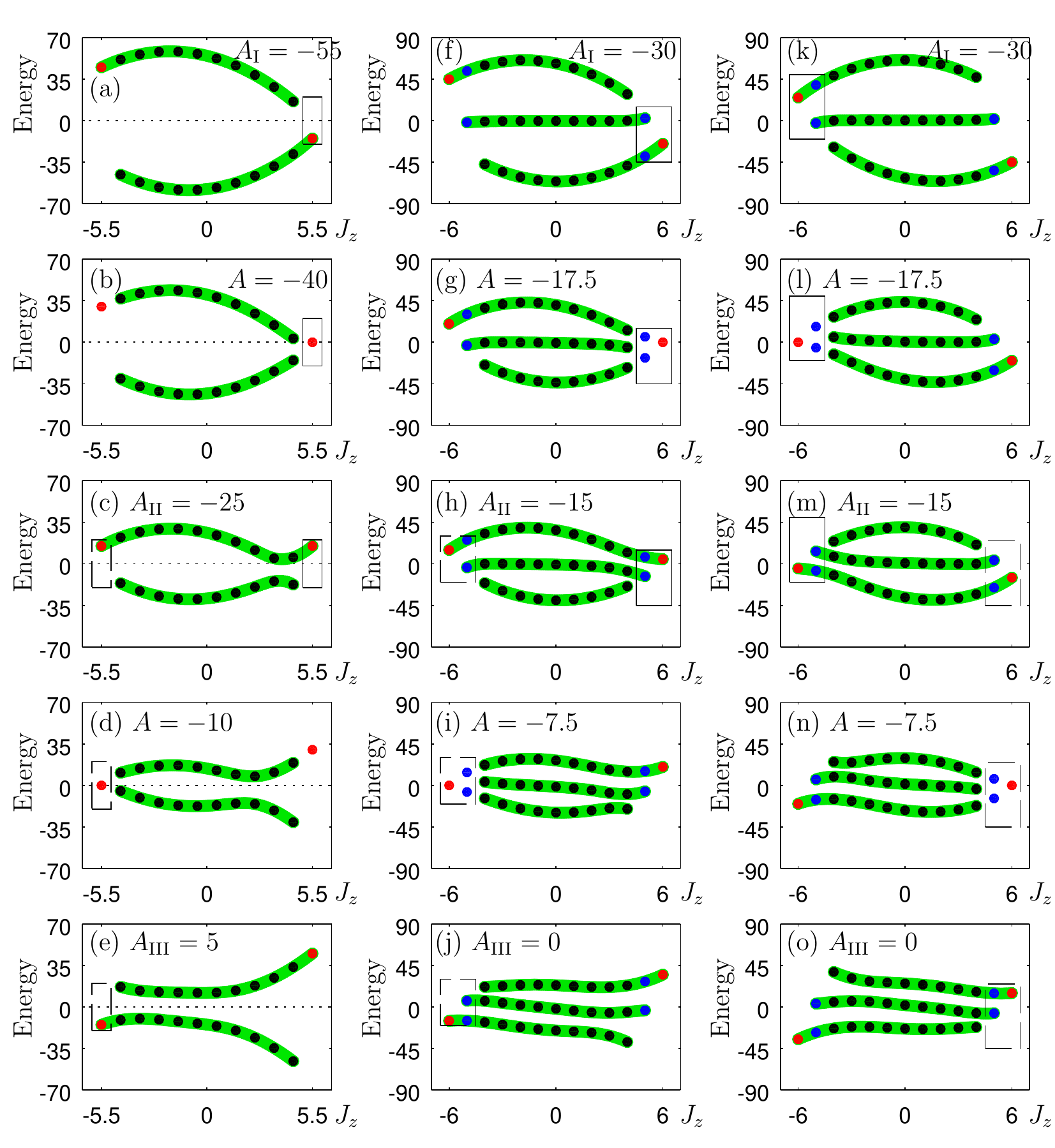}
\caption{Alternative representations of the energy level patterns shown in Figs.~\ref{Z2axialDelta}(a) and~\ref{FigG34}(a). Discrete points are the numerical solutions for the exact quantum Hamiltonian: the Hamiltonian commutes with the $J_z$ operator, so the former can be block diagonalized, where each block has a label given by a value of the projection of the total angular momentum on the $z$ axis. Then, each of these blocks is numerically diagonalized to obtain the discrete joint spectrum. The energy levels in a green region are assigned to the same band. The rectangles with full lines contain the edge states involved in the f\/irst iso-Chern domain crossing. The rectangles with dashed lines contain the edge states involved in the second iso-Chern domain crossing. Left column: evolution of the joint spectrum of the Hamiltonian~(\ref{L+L-}) with $S=\frac{1}{2}$, $L=5$, $\gamma=1+2i$, $d=1$, $\delta=3$. The dif\/ferent subf\/igures correspond respectively to dif\/ferent $A$ values with the $A=-40$ and $A=-10$ cases being associated respectively with the left and the right boundaries of domain~II in Fig.~\ref{Z2axialDelta}(a). Middle column: evolution of the joint spectrum of the Hamiltonian~(\ref{Hamiltonian_ref}) with $L=5$, $\gamma=1+2i$, $d=\frac{1}{2}$, $\delta=1$. The dif\/ferent subf\/igures correspond respectively to dif\/ferent $A$ values with the $A=-17.5$ and $A=-7.5$ cases being associated respectively with the left and the right boundaries of domain~II in Fig.~\ref{FigG34}. Right column: same as middle column, but with $\delta=-1$.} \label{figure_13}
\end{figure}

In order to see better the correspondence between the lattice of quantum states and the quantum results represented in Figs.~\ref{Z2axialDelta} and \ref{FigG34}, it is preferable to use an alternative presentation of the quantum results by plotting for several discrete values of control parameters the joint spectrum of two commuting observables, the energy $E$ and the integral of motion $J_z$. Applying this procedure with $S=\frac12$ and $S=1$ to the evolution of the classical energy-momentum map, we obtain Fig.~\ref{figure_13}, which shows that the rearrangement of the band structure (a set of levels located in the same green region def\/ines a~band) is done through the edge states $J_z=L+\frac{1}{2}$ (one level) and $J_z=-L-\frac{1}{2}$ (one level) for $S=\frac{1}{2}$ and $J_z=L+1$ (one level), $J_z=-L-1$ (one level), $J_z=L$ (two levels), and $J_z=-L$ (two levels) for $S=1$. For a f\/ixed value of the integral of motion $J_z$, the number of edge states is a linear function of $J_z$ while it is constant to $2S+1$ for the bulk states (see Fig.~\ref{DuistHFIG} for comparison). The part of the energy levels responsible for the reorganization is small in comparison to the total number of energy levels if $L\gg S$ and this assumption is quite natural as far as we treat $L_\alpha$ as classical variables whereas the number of quantum levels for the fast subsystem is supposed to be small.

The left column of Fig.~\ref{figure_13} shows joint spectra for $S=\frac{1}{2}$, $\delta>0$ and for f\/ive dif\/ferent values of the control parameter $A$, for exactly the same ef\/fective Hamiltonian that was used in Fig.~\ref{Z2axialDelta}(a). In fact, Figs.~\ref{figure_13}(a-e) and \ref{Z2axialDelta}(a) display in two dif\/ferent ways the same set of quantum energy levels originally living in the three-dimensional space of energy $E$, projection of the total angular momentum $J_z$, and control parameter~$A$. Fig.~\ref{Z2axialDelta}(a) gives the projection of this three-dimensional set on a $J_z=\mathrm{const}$ plane. Figs.~\ref{figure_13}(a--e) represent discrete sections of the three-dimensional pattern by $A=\mathrm{const}$ sections. Figs.~\ref{figure_13}(a,c,e) indicate that the band structures in domains~I, II, and III, calculated at $A_\mathrm{I}$, $A_\mathrm{II}$, and $A_\mathrm{III}$, are dif\/ferent. The band structure in Fig.~\ref{figure_13}(a) consists of two bands of $2L+1$ energy levels; the $J_z=L+\frac{1}{2}$ edge state belongs to the lower band, the $J_z=L-\frac{1}{2}$ edge state belongs to the upper band. The position of the edge states is exchanged in Fig.~\ref{figure_13}(e). In Fig.~\ref{figure_13}(c), the lower band has $2L$ levels (bulk states only) and the upper band has $2L+2$ levels (bulk states and the two edge states). Figs.~\ref{figure_13}(b,d) are associated with a~topological phase transition where the edge states cannot be assigned to a particular band.

The middle column of Fig.~\ref{figure_13} is associated with $S=1$, $\delta>0$. We clearly see the three dif\/ferent band structures at $A_\mathrm{I}=-30$ (domain~I), $A_\mathrm{II}=-17.5$ (domain~II) and $A_\mathrm{III}=0$ (domain~III). At $A_\mathrm{I}=-30$, we distinguish three distinct bands of $2L+1$ levels each. The $J_z=L+1$, $L$, $L$, $-L$, $-L$, and $-L-1$ levels respectively belong to: the lower band, the lower and middle band, the middle and upper band, and the upper band. When $A$ increases, the $J_z=L+1$ and $J_z=L$ edge states in the small rectangle in the right part of panels~(f),~(g) and~(h) move upwards. The band structure in the intermediate domain~II can be seen at $A=-15.0$. Due to the displacement of the $J_z=L+1$ and $J_z=L$ edge states, we see three bands but the number of quantum energy levels in each band has changed with respect to the situation in domain~I: two edge states left the lower band which has now only $2L-1$ levels. One edge state left the middle band to join the upper band and an other one left the lower band to join the middle band so the middle band has still $2L+1$ levels. The upper band gained two levels and has $2L+3$ levels. In panel (g), it becomes impossible to state to which bands belong the edge states and the bands only contain the bulk states. This redistribution of energy levels among the bands is correlated to a triple degeneracy in the semi-quantum model. Another redistribution is shown in the dashed rectangle in the left part of panels (h), (i) and~(j), where the $J_z=-L-1$ and $J_z=-L$ edge states are going downwards. The association of the edge states to the bands is not possible. Finally, at $A_\mathrm{III}=0$, there is again three well-def\/ined bands, each band with $2L+1$ quantum energy levels. The $J_z=L+1$, $L$, $L$, $-L$, $-L$, and $-L-1$ levels now respectively belong to: the upper band, the middle and upper band, the lower and middle band, and the lower band. This series of f\/igures clearly conf\/irm that the edge states are responsible for the redistribution of the energy levels between bands.

The right column of Fig.~\ref{figure_13} is again described with $S=1$ and the same Hamiltonian parameters as in the middle column, except that the sign of the $\delta$ parameter is reversed. We see that the band structures for $A\rightarrow -\infty$ and $A\rightarrow +\infty$ do not change. The main dif\/ference occurs in the domain~II: we see that the lower, middle and upper bands have respectively $2L+3$, $2L+1$ and $2L-1$ levels. In particular, the comparison between Fig.~\ref{figure_13}(h) and Fig.~\ref{figure_13}(m) shows us that the edge energy levels with $J_z=L+1$ and $L$ have shifted upward for $\delta >0$ but the edge state levels with $J_z=-L-1$, $-L$ have shifted downward for $\delta<0$. The comparison between Fig.~\ref{figure_13}(j) and Fig.~\ref{figure_13}(o) indicates that the energy levels with $J_z=-L-1$ and $-L$ have shifted downward for $\delta >0$ but the energy levels with $J_z=L+1$, $L$ have shifted upward for $\delta<0$.

The comparison of the redistribution of the energy levels between bands for the quantum problem with the qualitative modif\/ication of the lattice of quantum states in the image of the energy-momentum map leads us to the following qualitative picture of band rearrangement. If quantum states form a few number of energy bands (separated by gaps) with some initial distribution of energy levels between bands, the variation in the control parameter modif\/ies the quantum states, and accordingly the set of all energy levels can be split into two subsets, one of which is the subset of bulk states and the other is the subset of edge states. The bulk states belong to the same band during the variation of the control parameter and the edge states go from one band to another. In the transition, the edge state(s) belong equally to two bands and have properties qualitatively dif\/ferent from the bulk states of each band.

For the complete classical problem, a qualitative modif\/ication of the image of the energy-momentum map manifests itself as a modif\/ication of the system of critical values. In our example, the critical values take dif\/ferent positions on the boundary in the limit case as $|A|\to \infty$ and in an intermediate situation, some of the critical values appear as the isolated monodromy points inside the regular domain of the energy-momentum map. From the point of view of quantum interpretation, the presence of monodromy points is associated with a~``transition'' state between two qualitatively dif\/ferent band structures. From a complete classical perspective, the ``transition'' regime is another qualitatively dif\/ferent dynamical regime.

In the rest of this section, we make a remark on the lattice of quantum states. It is known that the elementary Hamiltonian monodromy of an integrable classical problem manifests itself as a lattice defect for the corresponding quantum problem~\cite{Ngoc, TopSolSt}.

\begin{figure}[t]\centering
 \includegraphics[scale=0.85]{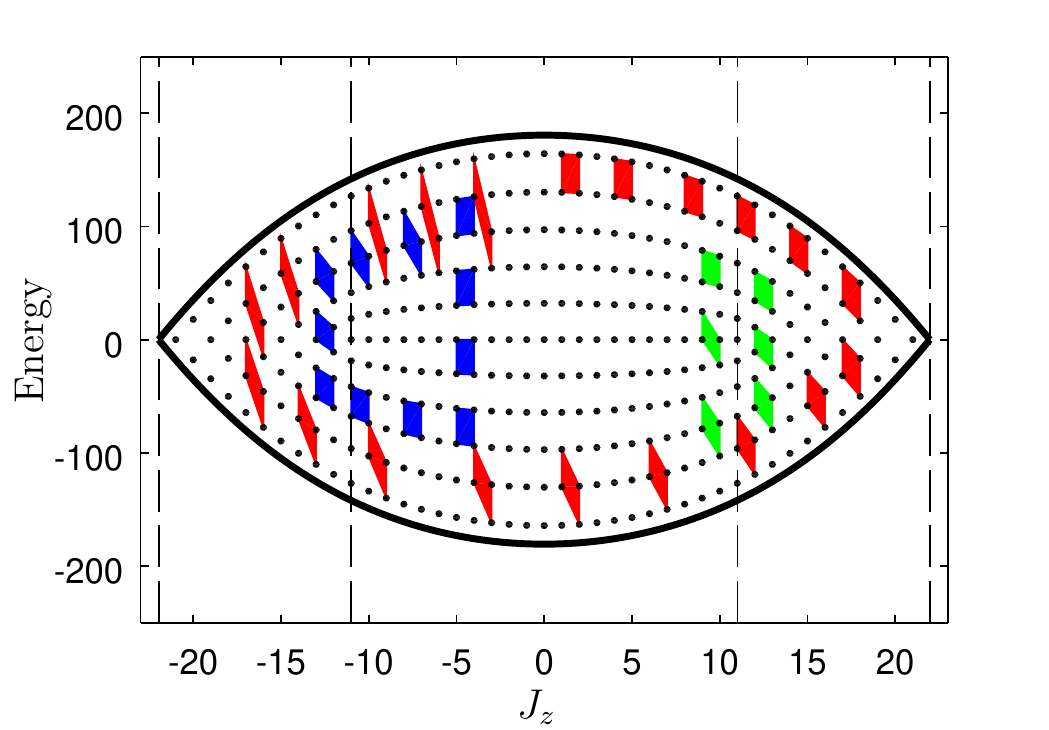}
\caption{Numerically computed joint spectrum of the quantum states superposed on the image of the energy-momentum map for Hamiltonian~(\ref{Hclass}) with $A=0=\delta=d=0$, $\gamma=1$, $S=5$, $L=16$. The boundary of the image of the energy-momentum map is calculated purely classically. Two elementary monodromy defects of the quantum state lattice become visible by following the evolution of an elementary cell of the lattice along a path surrounding each elementary defect (green cells and blue cells). Evolution of an elementary cell along a path surrounding both defects (red cells) shows that the two monodromy points have the same sign.}\label{ModelCells}
\end{figure}

Fig.~\ref{ModelCells} shows the lattice of quantum states done for a Hamiltonian with $\delta=0$, $d=0$, $\gamma=1$, $A=0$, and $S=5$. With such a choice of
parameters, the defects associated with two critical values of the energy-momentum map are located inside the regular domain and the evolution of elementary cells along paths is clearly visible. As is described in~\cite{PhysLettMono}, the quantum monodromy matrix is linked to a~change in the organization of the lattice of quantum states, which itself seems to be related to the redistribution phenomenon of energy levels.
Fig.~\ref{ModelCells} shows that the evolution of an elementary quantum cell around each of the two defects of the quantum state lattice (corresponding to the critical values of the classical energy-momentum map and to Hamiltonian monodromy) leads in an appropriately chosen basis to the same quantum monodromy matrix, $\left( \begin{smallmatrix} 1&0\\ -1&1\end{smallmatrix} \right) $. At the same time the evolution of the elementary cell along a~contour surrounding both elementary defects leads to the monodromy matrix $\left( \begin{smallmatrix} 1&0\\ -2&1\end{smallmatrix} \right) $. This demonstrates the additivity of several mono\-dro\-my points with the same vanishing cycle~\cite{Matveev}, while the monodromy matrix pos\-ses\-sing the of\/f-diagonal element of the opposite sign in the same basis corresponds to completely dif\/ferent defects composed of at least eleven elementary monodromy defects~\cite{CushmanZh, TopSolSt}.

We note, however, that the modif\/ication of the Chern numbers for the eigenline bundles in the case of the two-band model has dif\/ferent sign for the two degeneracy points associated with the two monodromy points of the completely classical model. The sign of the local delta-Chern contribution is presumably related to the direction of displacement of the critical value on the image of the energy-momentum map rather than with the fact itself of existence of an isolated critical value.

\section[Spectral f\/low and topological ef\/fects in molecular energy patterns]{Spectral f\/low and topological ef\/fects\\ in molecular energy patterns}\label{S_6}

\subsection{Spectral f\/low in a system of bands}

The qualitative rearrangement of the energy bands in a quantum molecular system under the variation of a control parameter can be formalized by introducing the notion of spectral f\/low. Originally, the spectral f\/low of a one-parameter family of self-adjoint operators is def\/ined to be the net number of eigenvalues passing through zero in the positive direction as the parameter runs~\cite{Atiyah}. This def\/inition is well-suited in the case of two bands. The extension of the concept to a system of $N$-bands, $N>2$, rather focuses on the local change of number of quantum states in each band, which we discuss below.

The spectral f\/low is naturally discussed in the quantum mechanical setting where the quantum levels of the Hamiltonian can be partitioned into bulk states that do not change band and edge states that redistribute among the bands along the variation of the control parameter. The assignment of the energy levels in term of bands is trivial for our Hamiltonian~(\ref{Hamiltonian_ref}) in the limits $\left|A\right|\rightarrow \infty$. The situation for intermediate $A$ values is less clear and the analysis of the redistribution of the quantum levels from the pure quantum point of view is made easier by introducing semi-quantum and classical arguments as in Section~\ref{S_5}.

Typically, the behavior of the eigenvalues of the semi-quantum Hamiltonian~(\ref{sqh}) suggests a~decomposition of the control parameter space of the Hamiltonian in iso-Chern domains (where the semi-quantum eigenvalues are non-degenerate), each corresponding to a dif\/ferent band structure, separated by walls where the semi-quantum eigenvalues are degenerate~\cite{AnnPhysIZh,IwaiZhWallCross,IwaiZhTheorChemAcc,I_Z}. Let us pick a representative point $A_i$ in each iso-Chern domain. For a~given control parameter $A_i$, each quantum level, bulk or edge state, is assigned to one and only one band. We suppose furthermore that the $A_i$'s are ordered, $\cdots<A_{i-1}<A_i<A_{i+1}<\cdots$ such that there is only one wall between the iso-Chern domain associated with $A_i$ and the one related to $A_{i+1}$. We give here def\/initions that are suitable when there is a redistribution of one or more eigenvalues among several bands upon the modif\/ication of the control parameter. Let the bands be numbered from bottom to top according to their energy: $b=0$ is the band with the lowest energy, $b=2S$ is the band of highest energy. When the control parameter increases from $A_i$ to $A_{i+1}$, $n$ quantum energy levels leaving the band $j$ and joining the band $k$ contribute to the local redistribution $f_{j \rightarrow k}([A_i,A_{i+1}])$ from band $j$ to band $k$ between $A_i$ and $A_{i+1}$:
\begin{gather*}
 f_{j\rightarrow k} ( [A_i,A_{i+1}] ) = +n.
\end{gather*}
This quantity is always positive by def\/inition.

\begin{definition}
 The \emph{local spectral flow} between $A_i$ and $A_{i+1}$ is the
 $2S+1$ component vector
\begin{gather*}
 \left( \begin{matrix}
 \Delta N_{2S} ( [A_i,A_{i+1} ] ) \\
 \vdots \\
 \Delta N_0 ( [A_i,A_{i+1} ] )
 \end{matrix} \right),
 \end{gather*}
of the local changes $\Delta N_{b}([A_i,A_{i+1}])$, i.e., the dif\/ferences between all the local redistributions joining the band~$b$ and all those local redistributions leaving the band $b$ between $A_i$ and $A_{i+1}$:
\begin{gather*}
 \Delta N_{b}([A_i,A_{i+1}]) = \sum_{k\neq b} \big\{ f_{k\rightarrow b}([A_i,A_{i+1}]) - f_{b\rightarrow k}([A_i,A_{i+1}]) \big\}.
\end{gather*}
The components of the local spectral f\/low can be negative, zero, or positive integers.
\end{definition}

\begin{definition}
The \emph{global spectral flow} of quantum energy levels is the cumulated sum of the local spectral f\/lows:
\begin{gather*}
 \left( \begin{matrix}
 \Delta N_{2S} \\
 \vdots \\
 \Delta N_0
 \end{matrix}\right)
 =\sum_i
\left( \begin{matrix}
 \Delta N_{2S}([A_i,A_{i+1}]) \\
 \vdots \\
 \Delta N_0([A_i,A_{i+1}])
\end{matrix}\right).
\end{gather*}
The components of the global spectral f\/low can be negative, zero, or positive integers.
\end{definition}

\subsection{Two bands}

The $A$-dependent Hamiltonian~(\ref{L+L-}) with $\delta > 0$ has edge state quantum energy levels responsible for the band rearrangement, to which we can assign local spectral f\/lows. Let $A_\mathrm{I}$, $A_\mathrm{II}$ and $A_\mathrm{III}$ be three values of the control parameter respectively chosen in domains I, II, and III, see Fig.~\ref{Z2axialDelta}. There is only one quantum energy level changing bands between $A_\mathrm{I}$ and $A_\mathrm{II}$. The local redistribution from the lower band $b=0$ to the upper band $b=1$, $f_{0 \rightarrow 1} ( [A_\mathrm{I},A_\mathrm{II} ] )=1$, is assigned to the edge state crossing the $E=0$ level at $A=-dL^2-\delta L$ and the local redistribution from the upper band to the lower band $f_{1\rightarrow 0} ( [A_\mathrm{II},A_\mathrm{III} ] )=1$ to the edge state crossing the $E=0$ level at $A=-dL^2+\delta L$.

The number of levels of the upper and the lower bands in the intermediate region~II of the control parameter $-dL^2- |\delta |L < A < -dL^2+ |\delta | L$ depends on the sign of $\delta$. If $\delta$ is positive as in Fig.~\ref{Z2axialDelta}, the upper band has $(2L+1)+1$ level, and the lower band has $(2L+1)-1$ level. If $\delta$ is negative, the numbers of quantum energy levels in each band are exchanged. If we compare the number of levels between the $A=\pm\infty$ limits, we see that each band gains one level and loses one level and the net result is no change in the number of quantum energy levels.

The Hamiltonian~(\ref{L+L-}) with $\delta = 0$ is a special case: the intermediate region~II is reduced to zero. There is a simultaneous redistribution of the two quantum energy levels in opposite directions which are synchronized due to the pseudo-symmetry of the Hamiltonian: $f_{0\rightarrow 1}\left(\left[A_\mathrm{I},A_\mathrm{III}\right]\right)=1$, $f_{1\rightarrow 0}\left(\left[A_\mathrm{I},A_\mathrm{III}\right]\right)=1$.

For the two-level semi-quantum Hamiltonian, Fig.~\ref{SchemChern2} showing the evolution of Chern numbers against the control parameter is in marked correspondence with the local spectral f\/low mentioned above. In addition, both of these qualitative modif\/ications occurring in the quantum and the semi-quantum models correspond to qualitative modif\/ications of the image of the energy-momentum map for the complete classical version of the same problem. As far as the observed modif\/ications are topological for classical and semi-quantum versions, they cannot be removed by a small deformation of Hamiltonian neither in classical nor in quantum picture.

An important feature of the analyzed qualitative phenomenon characteristic of systems with the energy-ref\/lection symmetry is the special organization of the observed modif\/ications. In fact, we observe two local qualitative modif\/ications which are generic for systems without time-reversal invariance, one at the north pole of the sphere, the other one at the south pole. These two phenomena occur simultaneously because of additional pseudo-symmetry (or energy-ref\/lection symmetry) of the Hamiltonian. Breaking this pseudo-symmetry (adding for example the term~$\delta L_z$) results in the splitting of this non-local phenomenon into two local ones occurring for dif\/ferent values of the control parameter.

\subsection[$N$-bands]{$\boldsymbol{N}$-bands}

In the semi-quantum model, the redistribution phenomenon is related with degeneracy among the eigenvalues of the Hamiltonian. When the degeneracy is generic, two and only two semi-quantum eigenvalues are degenerate at a particular point of the base space. In such a case, the quantum energy levels leaving one band join a neighboring band. When the degeneracy is multiple, it may happen that a quantum level leaving one band joins a band that is farther than its neighboring ones. For example, the quantum energy levels plotted in red in Fig.~\ref{FigG34} change bands by steps of two.

Let us observe from Fig.~\ref{FigG34}(a) with $S=1$, $\delta>0$, that there is one energy level (upward red line) that leaves the bottom band to join the top band. This level is represented by the $J_z=L+1$ red dot moving upwards in Figs.~\ref{figure_13}(f--j). In the semi-quantum model, the three eigenvalues of the $3\times 3$ matrix are degenerate at the north pole of the base space (a~sphere). Fig.~\ref{FigG34}(b) is an interpretation of the global change of bands between the $A=-\infty$ and $A=+\infty$ limits when only the levels going upward are considered. It is remarkable that this f\/igure has a local interpretation too, between $A_\mathrm{I}$ and $A_\mathrm{II}$. Section~\ref{S_5} discussed the problem of the redistribution of energy levels between bands by looking at the evolution with the control parameter $A$ of the lattice of quantum states in the energy-momentum map, see Figs.~\ref{figure_13}(f--j). Looking at the evolution of the edge states in Figs.~\ref{figure_13}(f--h), we understand that the transition of one energy level from the lowest band to the highest band between $A_\mathrm{I}$ and $A_\mathrm{II}$ must be accompanied by the transitions of two other levels: one level goes from the lower band to the middle band and one level goes from the middle band to the upper band. These two levels are the levels in blue in Fig.~\ref{FigG34}(a,b), going upwards with increasing $A$ and the blue $J_z=L$ dots in Figs.~\ref{figure_13}(f--h). As a~consequence, the local redistributions between $A_\mathrm{I}$ and $A_\mathrm{II}$ for $\delta>0$ are:
\begin{gather*}
 f_{0\rightarrow 2}([A_\mathrm{I},A_\mathrm{II}])=1,\qquad
 f_{0\rightarrow 1}([A_\mathrm{I},A_\mathrm{II}])=f_{1\rightarrow 2}([A_\mathrm{I},A_\mathrm{II}])=1.
\end{gather*}
The local spectral f\/low between $A_\mathrm{I}$ and $A_\mathrm{II}$ is then equal to:
\begin{gather*}
 \left( \begin{matrix}
 \Delta N_{2}([A_\mathrm{I},A_\mathrm{II}]) \\
 \Delta N_{1}([A_\mathrm{I},A_\mathrm{II}]) \\
 \Delta N_{0}([A_\mathrm{I},A_\mathrm{II}])
 \end{matrix} \right)
 =
 \left( \begin{matrix}
 f_{0\rightarrow 2}([A_\mathrm{I},A_\mathrm{II}])+f_{1\rightarrow 2}([A_\mathrm{I},A_\mathrm{II}]) \\
 f_{0\rightarrow 1}([A_\mathrm{I},A_\mathrm{II}])-f_{1\rightarrow 2}([A_\mathrm{I},A_\mathrm{II}]) \\
 -f_{0\rightarrow 2}([A_\mathrm{I},A_\mathrm{II}])-f_{0\rightarrow 1}([A_\mathrm{I},A_\mathrm{II}])
 \end{matrix} \right)
 =
 \left( \begin{matrix}
 2 \\ 0 \\ -2
 \end{matrix} \right),
\end{gather*}
consistent with the three-component vector of delta-Chern numbers $(-2,0,+2)^T$, see relations~(\ref{N_Chern_relation}) and~(\ref{chern_changes}) for $S=1$.

There is another triple degeneracy point in the semi-quantum model on the south pole for the control parameter separating the iso-Chern domains II (intermediate domain) and III (large values of $A$). Fig.~\ref{FigG34}(c) can be given two interpretations in a way similar to Fig.~\ref{FigG34}(b). First, it gives the global redistribution of the levels moving downward between the $A\rightarrow -\infty$ limit and the $A\rightarrow +\infty$ limit. It has a local interpretation between $A_\mathrm{II}$ and $A_\mathrm{III}$, too. Fig.~\ref{FigG34}(a) and the dashed rectangles in Figs.~\ref{figure_13}(h--j) indicates that three levels are going downward between $A_\mathrm{II}$ and $A_\mathrm{III}$ for $\delta>0$: one level in red travels from the upper band to the lower band and two other levels in blue shift downward by just one band:
\begin{gather*}
 f_{2\rightarrow 0}([A_\mathrm{II},A_\mathrm{III}])=1,\qquad f_{1\rightarrow 0}([A_\mathrm{II},A_\mathrm{III}])=f_{2\rightarrow 1}([A_\mathrm{II},A_\mathrm{III}])=1.
\end{gather*}
Between $A_\mathrm{II}$ and $A_\mathrm{III}$, the upper band has lost two quantum energy levels and the lower band has gained two levels. The middle band is left and joined by one level, implying no change in the number of energy levels in the middle band. If we compare the three bands for large negative values of $A$ and for large positive values of $A$, we can compute the global spectral f\/low:
\begin{gather*}
 \left( \begin{matrix}
 \Delta N_2 \\
 \Delta N_1 \\
 \Delta N_0
 \end{matrix} \right) =
 \left( \begin{matrix}
 \Delta N_{2}([A_\mathrm{I},A_\mathrm{II}])+\Delta N_{2}([A_\mathrm{II},A_\mathrm{III}]) \\
 \Delta N_{1}([A_\mathrm{I},A_\mathrm{II}])+\Delta N_{1}([A_\mathrm{II},A_\mathrm{III}]) \\
 \Delta N_{0}([A_\mathrm{I},A_\mathrm{II}])+\Delta N_{0}([A_\mathrm{II},A_\mathrm{III}])
 \end{matrix} \right) =
 \left( \begin{matrix}
 0 \\ 0 \\ 0
 \end{matrix} \right).
\end{gather*}
Globally there is no change in the number of energy levels in each band. The three-component vector of Chern numbers is $(0,0,0)^T$, and is again consistent with relations~(\ref{N_Chern_relation}) and~(\ref{chern_changes}).

The spectral f\/low mentioned above can be extended to the case of the band inversion for the $N$-bands described by the ef\/fective Hamiltonian~(\ref{Hamiltonian_ref}). In the general case for $N$-bands and $\delta>0$, there is again three iso-Chern domains. The f\/irst redistribution of energy levels occurs between $A_\mathrm{I}$ and $A_\mathrm{II}$, where we see upward local redistributions $f_{i\rightarrow j}([A_\mathrm{I},A_\mathrm{II}])=1$, $j>i$ only. There is $\sum\limits_{b=0}^{2S} (2S-b)=S(2S+1)$ such contributions between $A_\mathrm{I}$ and $A_\mathrm{II}$. Then the local change for band $b$ is equal to:
\begin{gather*}
 \Delta N_{b}([A_\mathrm{I},A_\mathrm{II}])=\sum_{i=0}^{b-1} f_{i \rightarrow b}([A_\mathrm{I},A_\mathrm{II}]) -\sum_{i=b+1}^{2S} f_{b \rightarrow i}([A_\mathrm{I},A_\mathrm{II}]) =-2 (S-b ).
\end{gather*}
The redistribution of the edge states between $A_\mathrm{I}$ and $A_\mathrm{II}$ is summarized in Table~\ref{summary_1}. The local spectral f\/low is simply the next to last column, from which the delta-Chern contributions (last column) can be easily deduced (sign change).

\begin{table}[t]\centering
 \caption{Redistribution of the edge states of Hamiltonian~(\ref{Hamiltonian_ref}) between $A_\mathrm{I}$ and $A_\mathrm{II}$ for $\delta>0$.\label{summary_1}}
 \begin{tabular}{ccccccc}
 \hline \hline
 band $b$ & going& coming & going & coming & $\Delta N_b([A_\mathrm{I},A_\mathrm{II}])$ & delta-Chern \\
 & upward & upward & downward & downward & &\\
 & from & from & from & from & &\\
 \hline
 $2S$ & $0$ & $2S$ & $0$ & $0$ & $2S$ & $-2S$ \\
 $2S-1$ & $1$ & $2S-1$ & $0$ & $0$ & $2\left(S-1\right)$ & $-2\left(S-1\right)$ \\
 $\vdots$ & $\vdots$ & $\vdots$ & $\vdots$ & $\vdots$ & $\vdots$ & $\vdots$ \\
 $b$ & $2S-b$ & $b$ & $0$ & $0$ & $-2\left(S-b\right)$ & $2\left(S-b\right)$ \\
 $\vdots$ & $\vdots$ & $\vdots$ & $\vdots$ & $\vdots$ & $\vdots$ & $\vdots$ \\
 $1$ & $2S-1$ & $1$ & $0$ & $0$ & $-2\left(S-1\right)$ & $2\left(S-1\right)$ \\
 $0$ & $2S$ & $0$ & $0$ & $0$ & $-2S$ & $2S$ \\
 \hline \hline
 \end{tabular}
\end{table}

The second change in bands occurs between $A_\mathrm{II}$ and
$A_\mathrm{III}$, with $S(2S+1)$ downward local
redistributions $f_{i\rightarrow
 j}([A_\mathrm{II},A_\mathrm{III}])=1$, $j<i$
only and local changes:
\begin{gather*}
 \Delta N_{b}([A_\mathrm{II},A_\mathrm{III}])=\sum_{i=b+1}^{2S} f_{i \rightarrow b}([A_\mathrm{II},A_\mathrm{III}]) -\sum_{i=0}^{b-1} f_{b \rightarrow i}([A_\mathrm{II},A_\mathrm{III}])
 =2\left(S-b\right).
\end{gather*}
The redistribution of the edge states between $A_\mathrm{II}$ and $A_\mathrm{III}$ is summarized in Table~\ref{summary_2}. As for Table~\ref{summary_1}, the local spectral f\/low and the delta-Chern contributions can respectively be found in the next to last and the last columns.

\begin{table}[t]\centering
 \caption{Redistribution of the edge states of Hamiltonian~(\ref{Hamiltonian_ref}) between $A_\mathrm{II}$ and $A_\mathrm{III}$ for $\delta>0$.\label{summary_2}}
 \begin{tabular}{ccccccc}
 \hline \hline
 band $b$ & going& coming & going & coming & $\Delta N_b([A_\mathrm{II},A_\mathrm{III}])$ & delta-Chern \\
 & upward & upward & downward & downward & &\\
 & from & from & from & from & &\\
 \hline
 $2S$ & $0$ & $0$ & $2S$ & $0$ & $-2S$ & $2S$ \\
 $2S-1$ & $0$ & $0$ & $2S-1$ & $1$ & $-2\left(S-1\right)$ & $2\left(S-1\right)$ \\
 $\vdots$ & $\vdots$ & $\vdots$ & $\vdots$ & $\vdots$ & $\vdots$ & $\vdots$ \\
 $b$ & $0$ & $0$ & $b$ & $2S-b$ & $2\left(S-b\right)$ & $-2\left(S-b\right)$ \\
 $\vdots$ & $\vdots$ & $\vdots$ & $\vdots$ & $\vdots$ & $\vdots$ & $\vdots$ \\
 $1$ & $0$ & $0$ & $1$ & $2S-1$ & $2\left(S-1\right)$ & $-2\left(S-1\right)$ \\
 $0$ & $0$ & $0$ & $0$ & $2S$ & $2S$ & $-2S$ \\
 \hline \hline
 \end{tabular}
\end{table}

As a consequence, the global change $\Delta N_b=\Delta N_{b}([A_\mathrm{I},A_\mathrm{II}])+\Delta N_{b}([A_\mathrm{II},A_\mathrm{III}])$ over the parameter space is zero for any band. These results are consistent with relations~(\ref{N_Chern_relation}) and (\ref{chern_changes}).

Without reference to physical conditions, band rearrangement amounts to a partition problem for the total number, $(2S+1)(2L+1)$, of energy levels. In view of the fact that $\sum\limits_{k=0}^{2S}2(S-k)=0$, we arrange $(2S+1)(2L+1)$ to obtain
\begin{gather*}
 (2S+1)(2L+1)= \sum _{k=0}^{2S} (2L+1) -\sum_{k=0}^{2S}2(S-k) =\sum_{k=0}^{2S} [2(L-S+k)+1].
\end{gather*}
This is exactly the same as the relation among the dimensions of representation spaces for the Clebsch--Gordan formula, $V_S \otimes V_L \simeq V_{L-S} \oplus \cdots \oplus V_{L+S}$ with $S<L$. This conformity with the Clebsh--Gordan formula is viewed as a consequence of the fact that the Hamiltonian~(\ref{Hamiltonian_ref}) is a~variant of the spin-orbit coupling Hamiltonian.

Furthermore, on denoting the Chern numbers by $\mathrm{Ch}_k=-2(S-k)$ or by $\mathrm{Ch}_r^\prime=2(S-r)$ with $r=2S-k$, the above equation is rewritten as
\begin{gather*}
 (2S+1)(2L+1) = \sum_{k=0}^{2S} (2L+1+\mathrm{Ch}_k ) = \sum_{r=0}^{2S} (2L+1+\mathrm{Ch}_r^\prime ).
\end{gather*}
Each term of the right-hand side of this equation is a generalization of equation~(\ref{N_Chern_relation}). The idea of the relation between band rearrangement and Chern number dates back to an early paper~\cite{FaurePRL}.

\subsection{Symmetries}

In conclusion of this section, we make remarks on discrete symmetry related to the spectral f\/low and the Chern number. The pseudo-symmetry treated in equation~(\ref{TransfGen}) and in equation~(\ref{energy-reflection symmetry, SQ}) (or equivalently called the particle-hole symmetry in Dirac theory) implies that a local redistribution between $A_i$ and $A_{i+1}$ from band $i$ to band $j$ is compensated by a local redistribution in the same interval but from band $j$ to band $i$. As a consequence, the number of levels in the dif\/ferent bands is invariant and correspondingly the modif\/ication of Chern number is zero.

Being interested in another discrete symmetry, we take up the time-reversal symmetry, which is considered as a reality condition imposed on the ef\/fective Hamiltonian. Under the presence of the time-reversal symmetry, two local phenomena occurring at dif\/ferent points of base space give the same delta-Chern contribution under the variation of a control parameter and the same redistribution of energy levels between energy bands. This leads to the conclusion that the components of the local spectral f\/low and the modif\/ication of the Chern number should be even for time-reversal invariant problems~\cite{OmriGat}.

\section{Conclusion}\label{S_7}
We started by studying a generic phenomenon of the rearrangement of energy levels between two bands in the presence of the axial symmetry and the energy-ref\/lection symmetry and have reached the study of the band inversion phenomenon for an arbitrary number of energy bands, which is characterized globally by using the correlation between the initial and f\/inal system of bands without entering into details of rearrangement process (see Fig.~\ref{RearrangS}). The simulta\-neous analysis of the full quantum problem, its semi-quantum analog, and its complete classical version, demonstrates that there exist perfect analogies among them. In particular, as far as band rearrangement is concerned, the evolution of the lattice of quantum states of\/fers another point of view for the interpretation of the redistribution of energy levels between bands. This allows us to apply the notion of topological phase transition for the qualitative phenomenon of the redistribution of energy levels between bands and to relate these transitions with explicitly calculated topological invariants for semi-quantum and classical models. In contrast to other studies of topological ef\/fects in molecular systems related with the formation of conical intersections in conf\/iguration space~\cite{LonguetHiggHerzb, Yarkony} or in momentum space~\cite{Lifshitz, Volovik} our analysis is based on topological ef\/fects manifested in the classical phase space associated with the quantum problem under study within the semi-quantum description and looked upon as a qualitative modif\/ication of the dynamical behavior through the recoupling of dif\/ferent degrees of freedom and the reorganization of the energy band structure. At the same time, it is important to note the similarity of the considered ef\/fect of the energy band reorganization in f\/inite particle quantum systems with other physically quite dif\/ferent but mathematically close phenomena. The most direct relation is with topological insulators in solid state physics~\cite{HasanKane}. A crucial point in energy bands treated in topological insulators is that there exists an energy level that closes a band gap. This corresponds to the energy level redistribution or band rearrangement discussed in our theory. Although the concept itself of topological insulators has been developed after the appearance of the f\/irst paper on energy level redistribution between energy bands in molecules \cite{EurophysL}, even before that, there were a number of publications serving as a prerequisite to the formulation of the qualitative theo\-ry of molecular band rearrangements. Along with the quantum Hall ef\/fect and Berry's phase discussions \cite{wilczek,simon, Thouless} there appeared the topological ideas developed by S.P.~Novikov during the description of a non-relativistic electron in a magnetic f\/ield \cite{novikov}\footnote{For further development of this direction see~\cite{GMN}.} and by G.~Volovik et al.\ in their works on singularities associated with gap nodes in superf\/luid~$\mathrm{^3He}$~\cite{grinevich,Volovik}. Appearance of topological singularities in band structures of solids was discussed by Herring~\cite{herring} almost at the beginning of the quantum mechanics. Further implications to the existence of new topological phases were suggested, for example, in \cite{abrikosov} and studied from the point of view of a mathematically much more elaborated approach in \cite{horava}. It is quite interesting also to compare the description of the Adler--Bell--Jackiw chiral anomaly by Nielsen and Ninomiya made in 1983 \cite{nielsen} (see especially their Fig.~2) on the basis of their previous works in particle physics on possible development of fermion theories on a lattice \cite{nielsenI,nielsenII} with the f\/igures representing the contact of the conduction band and the valence band at a single point. Recent experimental study of Adler--Bell--Jackiw anomaly \cite{zhang} conf\/irms the universality of the topo\-lo\-gi\-cal mathematical models in dif\/ferent domains of physics. Authors hope that the present paper will stimulate interest of a rather large physical community in the applications of topological ideas to molecular physics problems.

A natural continuation of the present work is the extension of the present qualitative ana\-lysis to the redistribution of energy levels between energy bands of molecular quaternionic sys\-tems~\mbox{\cite{Avron3, Dyson}}, i.e., molecular systems with an half integer spin invariant under time-reversal, and comparison of the corresponding topological transitions with the physical description of the AII class of topological insulators and with the mathematical description of the unformal mathematical trinity between mathematical models over real, complex, and quaternionic coef\/f\/icients~\cite{MatTrinity, ArnoldFront, DeNittis}.

\vspace{-1mm}

\subsection*{Acknowledgements}
Part of this work was supported by a Grant-in Aid for Scientif\/ic Research No.~26400068 (T.I.) from JSPS.
\vspace{-1mm}

\pdfbookmark[1]{References}{ref}
\LastPageEnding

\end{document}